\documentclass[useAMS,usenatbib]{mn2e}
\usepackage{graphicx}
\usepackage{float}
\usepackage{subcaption}
\usepackage[usenames,dvipsnames]{xcolor}

\newcommand\hd[2]{\multicolumn{4}{c|}{\bfseries\begin{tabular}{@{}c@{}}#2\end{tabular}}}

\title[A global look at X-ray time lags]{A global look at X-ray time lags in Seyfert Galaxies}
\author[Kara et al.]{E. Kara$^{1,2}$\thanks{E-mail:
ekara@astro.umd.edu}, W. N. Alston$^{1}$, A. C. Fabian$^{1}$, E. M. Cackett$^{3}$, P. Uttley$^{4}$,
\newauthor
C. S. Reynolds$^{2}$ and A. Zoghbi$^{5}$ \\
$^{1}$Institute of Astronomy, Madingley Rd, Cambridge CB3 0HA\\
$^{2}$Department of Astronomy, University of Maryland, College Park, MD 20742-2421, USA\\
$^{3}$Department of Physics and Astronomy, Wayne State University, Detroit, MI 48201, USA\\
$^{4}$Astronmical Institute `Anton Pannekoek', University of Amsterdam, Postbus 94249, 1090 GE Amsterdam, the Netherlands\\
$^{5}$Department of Astronomy, University of Michigan, Ann Arbor, MI, 48109, USA\\
}

\begin{document}

\date{\today}

\pagerange{\pageref{firstpage}--\pageref{lastpage}} \pubyear{2016}

\maketitle

\label{firstpage}

\begin{abstract}
X-ray reverberation, where light-travel time delays map out the compact geometry around the inner accretion flow in supermassive black holes, has been discovered in several of the brightest, most variable and well-known Seyfert galaxies.  In this work, we expand the study of X-ray reverberation to all Seyfert galaxies in the {\em XMM-Newton} archive above a nominal rms variability and exposure level (a total of 43 sources).  $\sim$50~per cent of source exhibit iron~K reverberation, in that the broad iron~K emission line responds to rapid variability in the continuum.  We also find that on long timescales, the hard band emission lags behind the soft band emission in 85 per cent of sources.  This `low-frequency hard lag' is likely associated with the coronal emission, and so this result suggests that most sources with X-ray variability show intrinsic variability from the nuclear region.  We update the known iron~K lag amplitude vs. black hole mass relation, and find evidence that the height or extent of the coronal source (as inferred by the reverberation time delay) increases with mass accretion rate.

\end{abstract}


\section{Introduction}

In recent years, X-ray reverberation has opened a new way to investigate the inner accretion flow around supermassive black holes. {\em XMM-Newton} and {\em NuSTAR} observations of the high-frequency variability have shown that the soft excess \citep{fabian09,demarco13}, broad iron K line \citep{zoghbi12,kara13c} and Compton hump \citep{zoghbi14,kara15a} lag behind the continuum emission, suggesting light travel distances of a few gravitational radii. 
Previous studies have shown reverberation in the best individual cases, where the sources are highly variable, bright and have long observations.  Now that the phenomenon of iron~K lags has been established in many individual studies, it is important to determine its prevalence in the wider Seyfert population.  Therefore in this work we conduct a global look at the X-ray time lags in Seyfert galaxies observed with {\em XMM-Newton}.

X-ray reverberation lags were first robustly discovered in the source 1H~0707-495, which is one of the most highly variable and well-observed Seyfert galaxies in the {\em XMM-Newton} archive \citep{fabian09,zoghbi10}.  In that study, the authors showed that for the emission varying on short timescales ($\sim 1000$~s), the 0.3--1~keV band (which we will refer to as the {\em soft band}) lagged behind the continuum-dominated 1--4~keV band (the {\em hard band}) by 30 seconds.  This {\em soft lag} was interpreted as due to the light-travel distance between the continuum-emitting corona and the accretion disc that fluoresces after being irradiated by the continuum.  Several studies confirmed this discovery in other bright, well-known Seyferts (e.g. \citealt{zoghbi11b,emm11,demarco11,legg12}).  See \citet{uttley14} for a detailed review.

\citet{demarco13} performed the first systematic look at the frequency dependence of the lags between the soft and hard bands (roughly the 0.3--1~keV and 1--4~keV bands).  They performed this analysis on the CAIXA catalog \citep{bianchi09} of unobscured, radio-quiet Seyferts in the {\em XMM-Newton} archive.  They looked at the time lags for all sources in that catalog that have long ($>40$~ks) observations, have a fractional excess variance above zero at the $>2 \sigma$ confidence level \citep{ponti12}, and have published black hole mass estimates. In their sample of 32 sources, 15 sources showed a soft lag with significance $>97$~per cent.  They also found that the amplitude of the lag scaled with black hole mass.  Converting the amplitude of the soft lag to a light travel distance placed the corona at a height of between 1--10$r_{\mathrm{g}}$ for all 15 sources (where $r_{\mathrm{g}}=GM/c^{2}$).  

Robust confirmation of the interpretation of high-frequency soft lags came with the discovery of iron~K reverberation in the very bright Seyfert, NGC~4151 \citep{zoghbi12}.  In that work, the authors presented the energy-resolved time lags, and found that the continuum-dominated band varied before the line centroid.  Moreover, by examining the more rapid variability (presumably produced from emission closer to the black hole), they found a shorter time delay between the continuum and the gravitationally redshifted wing of the iron line.  {\em Iron~K reverberation} (defined as the delay of the iron line centroid emission with respect to continuum-dominated bands) has been found now in several sources \citep{kara13b, zoghbi13}, and most recently it has been found associated with the frequency of quasi-periodic oscillations in MS22549-3712 \citep{alston15}.

In \citet{kara13c} for the cases of Ark~564 and Mrk~335, we found that the high-frequency variability showed iron~K reverberation, while the low-frequency variability showed the time delay increasing steadily with energy.  This {\em low-frequency hard lag} has been observed for several decades, beginning with the discovery of hard lags in the BHB Cyg~X-1 \citep{miyamoto88} and later in the AGN NGC~7469 \citep{papadakis01}.  Low-frequency X-ray lags (and low-frequency X-ray variability, in general) are observed over many decades in timescale, including timescales that are orders of magnitude longer than the viscous timescale of the inner accretion flow.  \citet{lyubarskii97} proposed that the variability originates at a range of radii through fluctuations in the efficiency of the angular momentum transport, which in turn varies the mass accretion rate.  These fluctuations occur at a range of radii (and therefore a range of timescales) and propagate inwards on the viscous diffusion timescale. \citet{kotov01} and later \citet{arevalo06} extended this idea further to explain the time lags in the X-ray emission. These authors suggested that if these long timescale fluctuations modulate an inhomogeneous X-ray emitting corona that produces a softer X-ray spectrum at large radii and a harder spectrum at small radii, then these inwardly propagating fluctuations would cause the soft photons to respond first before the hard photons.  Within the theory of propagating fluctuations, there is no prediction for high-frequency soft band lags, and therefore the idea of a separate mechanism (reverberation) producing the high-frequency lags is consistent with the overall picture.

\begin{table*}
\centering
\begin{tabular}{l l l l l l l}
\hline
{\bf Name} & {\bf Seyfert Type} & {\bf  log(Mass) (log($M_{\odot}$))} & {\bf Reference} & {\bf $F_{var}$} & {\bf Total Counts} & {\bf  Exposure (s)} \\
\hline
1H 0707-495	&	1	&	6.31	&	1	&	0.527	&	1.44E+05	&	1.07E+06	\\
Ark 564	&	1	&	6.27	&	P	&	0.213	&	1.61E+06	&	5.76E+05	\\
ESO 113-G010	&	1.8	&	6.74	&	P	&	0.159	&	3.77E+04	&	8.91E+04	\\
ESO 362-G18$^{*}$	&	1.5	&	7.65	&	2	&	0.131	&	7.06E+04	&	6.07E+04	\\
ESO 511-G030	&	1	&	8.66	&	P	&	0.050	&	3.04E+05	&	1.16E+05	\\
IC 4329A	&	1.2	&	8.3	&	G	&	0.028	&	1.63E+06	&	1.33E+05	\\
IRAS 05078+1626	&	1.5	&	7.55	&	P	&	0.063	&	1.81E+05	&	5.84E+04	\\
IRAS 13224-3809	&	1	&	6.8	&	G	&	0.612	&	4.25E+04	&	5.04E+05	\\
IRAS 13349+2438	&	1	&	7.7	&	G	&	0.211	&	5.56E+04	&	1.40E+05	\\
IRAS 17020+4544	&	1	&	6.54	&	P	&	0.156	&	1.46E+05	&	1.63E+05	\\
IRAS 18325-5926$^{*}$&	2	&	6.4	&	3	&	0.215	&	8.43E+05	&	3.33E+05	\\
IZW1	&	1	&	7.4	&	G	&	0.150	&	6.06E+04	&	8.50E+04	\\
MCG-02-14-009	&	1	&	7.13	&	P	&	0.126	&	7.15E+04	&	1.25E+05	\\
MCG-5-23-16	&	1.9	&	7.92	&	5	&	0.074	&	4.80E+06	&	3.77E+05	\\
MCG-6-30-15	&	1	&	6.3	&	P	&	0.212	&	3.77E+06	&	6.99E+05	\\
Mrk 1040	&	1	&	7.6	&	G	&	0.081	&	3.13E+05	&	8.86E+04	\\
Mrk 205	&	1	&	8.32	&	P	&	0.075	&	1.82E+05	&	1.33E+05	\\
Mrk 335	&	1	&	7.23	&	R	&	0.177	&	4.28E+05	&	3.18E+05	\\
Mrk 586	&	1	&	7.6	&	G	&	0.248	&	1.25E+04	&	4.84E+04	\\
Mrk 704	&	1.2	&	8.11	&	P	&	0.248	&	1.25E+04	&	4.84E+04	\\
Mrk 766	&	1	&	6.822	&	R	&	0.228	&	1.19E+06	&	5.90E+05	\\
Mrk 841	&	1.5	&	8.52	&	P	&	0.142	&	5.73E+04	&	4.68E+04	\\
MS22549-3712	&	1.5	&	7	&	5	&	0.100	&	1.00E+05	&	1.00E+05	\\
NGC 1365$^{*}$	&	1.8	&	7.6	&	G	&	0.234	&	8.70E+05	&	9.48E+05	\\
NGC 3227	&	1.5	&	6.775	&	R	&	0.096	&	4.23E+05	&	1.03E+05	\\
NGC 3516	&	1.5	&	7.395	&	R	&	0.088	&	1.79E+06	&	4.65E+05	\\
NGC 3783	&	1.5	&	7.371	&	R	&	0.066	&	1.18E+06	&	1.98E+05	\\
NGC 4051	&	1	&	6.13	&	R	&	0.400	&	2.00E+05	&	2.00E+05	\\
NGC 4151	&	1.5	&	7.65	&	R	&	0.077	&	7.93E+05	&	1.45E+05	\\
NGC 4395	&	1.8	&	5.449	&	R	&	0.392	&	5.67E+04	&	9.83E+04	\\
NGC 4593	&	1	&	6.882	&	R	&	0.172	&	4.06E+05	&	7.60E+04	\\
NGC 4748	&	1	&	6.407	&	R	&	0.161	&	7.93E+04	&	6.53E+04	\\
NGC 5506$^{*}$	&	1.9	&	7.4	&	G	&	0.097	&	2.03E+06	&	1.79E+05	\\
NGC 5548	&	2	&	7.718	&	R	&	0.039	&	5.01E+05	&	9.55E+04	\\
NGC 6860	&	1.5	&	7.6	&	G	&	0.070	&	3.39E+05	&	1.20E+05	\\
NGC 7314	&	2	&	6.7	&	6	&	0.223	&	8.65E+05	&	3.53E+05	\\
NGC 7469	&	1.5	&	6.956	&	R	&	0.078	&	6.39E+05	&	1.63E+05	\\
PG 1211+143	&	1	&	7.61	&	P	&	0.118	&	9.22E+04	&	1.85E+05	\\
PG 1244+026 	&	1	&	7.26	&	7	&	0.190	&	5.71E+04	&	1.26E+05	\\
PKS 0558-504	&	1	&	7.8	&	8	&	0.154	&	1.15E+06	&	6.25E+05	\\
RE J1034+398	&	1	&	6.6	&	9	&	0.170	&	4.00E+05	&	4.00E+05	\\
RX J0136.9-3510	&	1	&	7.9	&	G	&	0.315	&	6.12E+03	&	5.35E+04	\\
SWIFT J2127.4+5654	&	1	&	7.18	&	10	&	0.137	&	1.48E+06	&	4.66E+05	\\
\hline
\end{tabular}
\caption[The sample]{Table of the 43 sources in this sample, including the name of the source, black hole mass, the reference on the black hole mass, the 2--10~keV fractional excess variance on 40~ks segments, the total 2--10 keV counts included in the analysis, and the total exposure length in seconds. To give some indication of whether sources are obscured or not, we note with an asterisk the sources that are not in the CAIXA catalog \citep{bianchi09}, which only contains unobscured sources with $N_{\mathrm{H}}<2\times 10^{22}$~cm$^{-2}$.  Note that PG~1244+026, NGC~4748 and SWIFT J2127.4+5654 are not in CAIXA because they were observed after this publication. We used estimates of black hole masses that were made independently of the X-ray properties. For the black hole mass references, `R' means Optical reverberation mass estimate (see text), `P' means the mass was taken from the Optical measurements in \citet{ponti12} (in order to have an estimate that was independent of the X-ray variability properties). `G' refers to mass estimates from \citet{gmv12} The numbered references refer to specific papers: (1) \citet{bian03}; (2) \citet{agis14}; (3) Iwasawa et al., {\em submitted}; (4) \citet{oliva95}; (5) \citet{alston15}; (6) \citet{schulz94}; (7) \citet{marconi08};  (8) \citet{papadakis05}; (9) \citet{alston14b}; (10) \citet{malizia08}.}
\label{obs_table}
\end{table*}


While low-frequency hard lags appear to be common in most sources with high-frequency soft lags (see e.g. \citealt{demarco13}), there are hints in some sources that a low-frequency soft lag can exist.   \citet{alston13} showed for NGC~4051, that during high-flux intervals of the observation, the typical low-frequency hard lag and high-frequency soft lag was present. However, the low-flux segments of the observation showed that the low-frequency hard lag disappeared, and became a low-frequency soft lag.  Another example of a low-frequency soft lag is in NGC~1365, which is a source that is known to have complex absorbers that can eclipse the intrinsic emission from the nucleus.  When the source was in a largely unobscured state, we found clear evidence for a high-frequency iron~K lag (and possibly even the Compton hump lag; \citealt{kara15a}).  However, at low frequencies, the soft band lagged behind the hard band.  This low-frequency soft lag was only present in this particular observation.  In that work, we suggested that the low-frequency soft lag was due to a decrease in the column density of the distant absorber during the observation (i.e. from a neutral cloud moving out of our line-of-sight).  In this case, the hard photons transmit first before the soft photons that can only escape later, as the source becomes less obscured.  We do not yet understand these low-frequency soft lags, and there is no consensus on whether these lags are due to fluctuations that are intrinsic to the inner accretion flow or are caused by more distant material.  One of the goals of this project is to understand how common this feature is, and under what circumstance it occurs.

In this paper, we would like to explore three main questions: (1) is iron~K reverberation common? (2) how ubiquitous is the low-frequency hard lag? and (3) how many sources show signs of low-frequency soft lags?  To answer these questions, we have performed a global look at the lag-energy spectra of all variable, well-observed sources in the {\em XMM-Newton} catalog.  This paper is organized as follows: in Section~\ref{sample_section}, we present the sample used for this study.  In Section~\ref{obs}, we briefly describe the data reduction and the Fourier analysis method.  In Section~\ref{results} we show results of the lag-frequency and frequency-dependent lag-energy spectra for the sample.  In the discussion in Section~\ref{discuss}, we will discuss the results in the context of our three main questions.

\section{The sample}
\label{sample_section}

In this systematic look at the X-ray time lags in Seyfert galaxies, we analyse archival observations taken with the {\em XMM-Newton} observatory that are public as of 2015 January 1.  This study is largely motivated by X-ray reverberation, but we are also interested in the ubiquity of the low-frequency lags and the possible effect of absorption on the lags, so the search is not limited to just `bare' (unobscured) Seyfert 1s, but rather is extended to all Seyferts, even those that have known neutral absorption.  

The only cuts made are in the length of the observation (they must have longer than 40~ks exposure) and in the variability.  In order to measure a lag, sources must have some variability, and so we make a cut based on the 0.3--10~keV unbinned periodogram.  We used the results of \citet{gmv12} where the authors present the periodograms of 104 AGN observed with {\em XMM-Newton}. We require that the lowest frequency bin (of the unbinned periodogram) have power that is at least two orders of magnitude above the level of the Poisson noise. We also require that at least 5 frequency bins have variability power above Poisson noise. This is a conservative cut that ensures that there are some features in the light curves, rather than just a constant flux, or a flux that increases or decreases steadily over the length of the observation.  These cuts allow us to capture a general picture of the time lags in many sources, even those that are not highly variable.  We calculate the average time lags for a particular source based on all observations that individually fit the criteria described above.    

\begin{figure}
\includegraphics[width=\columnwidth]{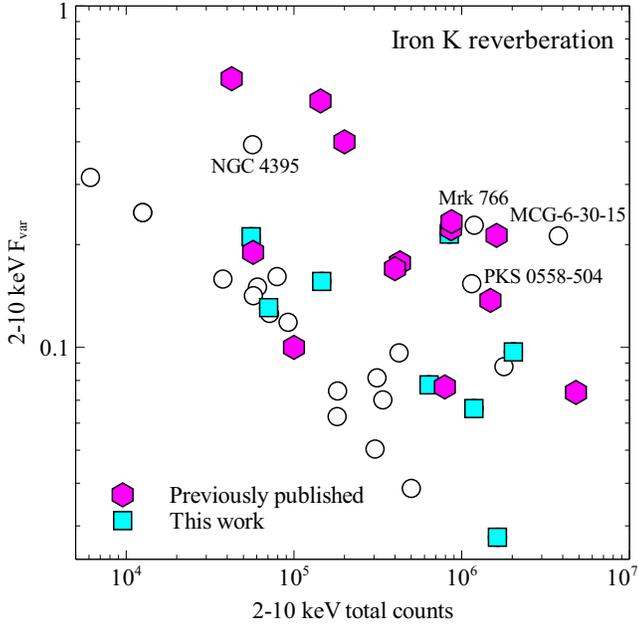}\\
\caption{The 2--10 keV fractional excess variance vs. the 2--10 keV total counts for all the sources in our sample (43 sources).  The sources in the top right portion have statistically the best data (i.e. highest level of variability, highest count rate and longest exposures).  Many sources in the {\em XMM-Newton} archive exist in the bottom-left portion of the plot, but these sources did not meet our selection criteria. The magenta hexagons show the 13 sources with previously published iron~K reverberation lags.  The cyan squares show the 8 sources found through this work to have iron~K reverberation at $>97$ per cent confidence (see text for more on measurement of statistical significance). The rest are shown as open circles. 20 out of 43 (or roughly half) of the sources in our sample show reverberation. }
\label{fvar_reverb}
\end{figure}

With these cuts our final sample consists of 43 Seyfert galaxies with varying flux, exposures and variability powers (see Table~\ref{obs_table}).  In Fig.~\ref{fvar_reverb} we plot the sample in the rms-counts plane (the distinction between filled markers and open circles will be discussed later).  We calculate the 2--10~keV fractional excess variance on 40~ks segments and 512~s time bins, using the prescription in \citet{vaughan03}. We plot this versus the total 2--10~keV counts (i.e. combining information on the exposure and the flux of the source).  The resulting rms-counts plane shows the cases with the highest quality data in the top-right portion of the plot.  Many sources in the {\em XMM-Newton} catalog lie on the bottom-left portion of the plot, but those sources have not been included in our sample because of their lack of variability and source counts.

\begin{figure}
\includegraphics[width=\columnwidth]{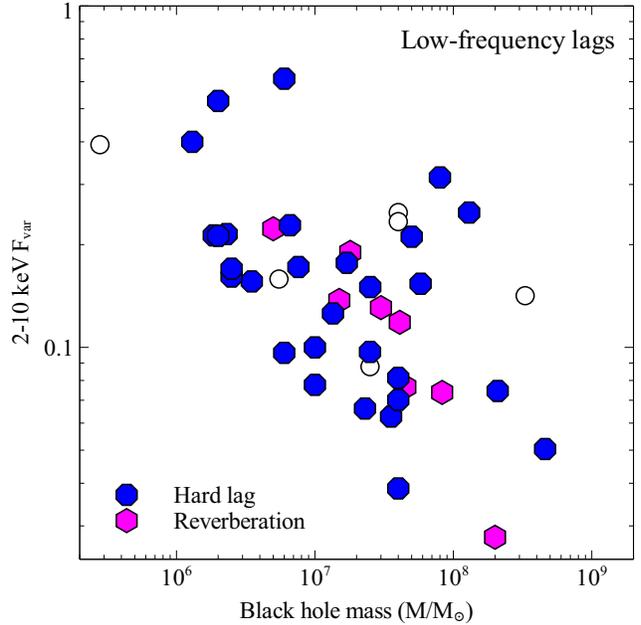}\\
\caption{The 2--10 keV fractional excess variance vs. black hole mass relation for our sample of 43 sources.  The blue octagons show the sources that have a hard lag that increase log-linearly with energy and do not show signs of iron~K reverberation at low frequencies (29 sources).  The magenta hexagons represent sources where an iron~K lag is seen at the lowest frequencies that we can probe.  Since we cannot probe lower frequencies, we cannot determine if these sources also have hard lags at lower frequencies (8 sources).  If we do not count these 8 sources, then we find that a hard lag is found at the lowest frequencies in 29/35 sources (or 83 per cent of sources). The sources in our sample that do not show either iron~K reverberation or hard lags at low-frequencies are shown as the unfilled circles. These sources are named in the text.  We emphasize that there are four sources which show the curious behaviour in which there is a hard lag that increases with energy, but at lower frequencies, this hard lag disappears or turns over to a soft lag. These four sources are MCG-6-30-15, NGC~3227, Mrk~1040, and Mrk~205.}
\label{fvar_hard}
\end{figure}

While the rms-counts plane is a good indicator for the statistical favorability of a source, there are some biases in this plot.  The value of the fractional excess variance is largely dependent on the frequency at which a source becomes dominated by Poisson noise.  Therefore, higher mass objects, which become dominated by Poisson noise at lower frequencies and show much of their variability on timescales longer than 40~ks, will have lower fractional excess variances.  This has been shown clearly through the well-known relation between rms and black hole mass (e.g. \citealt{ponti12,ludlam15}).  This relation shows that with {\em XMM-Newton} we are observationally biased to detecting lags in lower mass objects.  We show the known rms-mass correlation for our sample of 43 sources in Fig.~\ref{fvar_hard} (again, the distinction between colored and open markers will be discussed in the following sections).  For this analysis, Optical reverberation mass estimates were available for 12 sources in our sample.  They were collected from the public web database described in \citet{bentz15} using the default $<f>$ value of 4.3 from \citet{grier13}.  The references for the remaining black hole mass estimates are shown in Table~\ref{obs_table}. For all sources without reverberation mass estimates, we make a conservative estimate on the error of 0.5 dex.  These were estimated with various techniques, mostly the $R_{\mathrm{BLR}}-L$ relation (e.g. \citealt{kaspi00,bentz13}) or the velocity dispersion. 

Another slight bias in the rms-counts plane is due to neutral absorption, which causes less flux and variability in the soft band (primarily below 2~keV).  The sources that are partially covered by absorbing matter may not be useful for reverberation studies below 2 keV, but at high energies may reveal clear iron~K reverberation (e.g. NGC~4151 or NGC~1365).  We try to minimize this effect by plotting the 2--10~keV counts and excess variance, however some heavily absorbed sources may not appear close to the top-right corner in the rms-counts plane, but can still be used for reverberation.  

In Section~\ref{results}, we present the results of the time lag analysis for the 43 sources, but first we describe the data reduction and the Fourier method used to calculate the frequency-dependent time lags.

\begin{figure*}
\includegraphics[width=\textwidth]{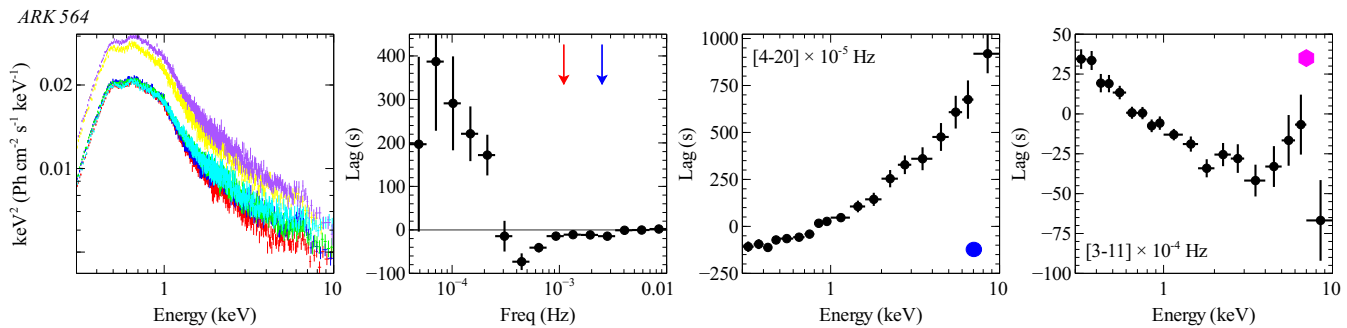}
\caption{An example of the results of this analysis for the case of Ark~564.  Similar plots for each source in Table~\ref{obs_table} can be found in the appendix.  The leftmost panel (a) shows the unfolded time-integrated fluxed energy spectrum of each observation that was included in the analysis. The next panel (b) shows the lag-frequency spectrum between the soft band (0.3--1~keV) and hard band (1--4~keV).  The blue and red arrows refer to the frequency at which the the PSD of the lowest and highest energy bin (in the lag-energy spectra) become dominated by Poisson noise.  The panels on the right (c,d) show the lag-energy spectra at particular frequencies.  The frequencies are determined by the lag-frequency spectra.  In the case of Ark~564, the frequencies used in panel (c) were selected because this is where a hard (positive) lag is observed, and the frequencies in panel (d) were selected based on where there is a soft (negative) lag in the lag-frequency spectrum.  The symbols in the corners of the lag-energy spectra indicate generally what catagory they fall into: magenta hexagon = previously published iron~K reverberation; cyan square = iron K reverberation found through this work; blue circle = low-frequency hard lag; red diamond = low-frequency soft lag. }
\label{ark564_lags}
\end{figure*}

\section{Observations and Method}
\label{obs}

\subsection{Data Reduction}

For this analysis, we use the data from the {\em XMM-Newton} EPIC-pn camera because of its higher effective area and fast readout \citep{struder01,jansen01}.  We reduced the data using the {\em XMM-Newton} Science Analysis System (SAS v. 14.0.0) and the newest calibration files. We started with the observation data files (ODFs) and followed the standard procedures.  The events were filtered with the conditions PATTERN$\le 4$ and FLAG==0.  The data were cleaned for high background flares (in which the background source counts were greater than half the source counts of that particular observation).  The source extraction regions are circular regions of radius 35 arcsec centered on the maximum source emission.  The background regions were also circular regions with a radius of at least 35 arcsecs.  For observations taken in Small Window Mode, this is the largest possible background region, but for observations taken in Large Window Mode or Full Frame Mode, we were able to use larger background regions. We used the tool {\sc epiclccorr} to produce background subtracted light curves in several energy bands. We perform the analysis on the full light curve (i.e. not on shorter segments), so that we can reach the lowest frequencies possible.

\subsection{The Fourier Method}

To compute the time lags, we used the Fourier technique outlined in \citet{nowak99}, and presented in detail in \citet{uttley14}.  Briefly, we take the Fourier transform of the light curves in two different energy bands.  We multiply the Fourier transform of one light curve and the complex conjugate of the Fourier transform of the other light curve.  This complex-numbered product is called the cross spectrum.  We bin the cross spectrum in equal logarithmic frequency bins in order to minimize the random scatter from individual frequencies and to visualize broader trends in the lags. The number of frequency bins is dictated by the the quality of the data (i.e. more bins for higher quality data). The general rule of thumb is that the lags in adjacent frequency bins should be $\sim 1 \sigma$ of each other.  The minimum frequency is defined by the inverse of the length of the observation (though we do not include the lowest frequency bin as this is often biased by red noise leakage), and the maximum frequency is set by the frequency at which the 0.3--10~keV PSD becomes dominated by Poisson noise. If there are multiple observations that fit our sample criteria, we take the average of the binned cross spectrum.  The argument of the cross spectrum is the phase lag between the two light curves.  The time lag, $\tau(f) = \phi(f)/2\pi f$, where $\phi(f)$ is the frequency-dependent phase and $f$ is the midpoint frequency of the frequency bin.  For this analysis, the lag-frequency spectrum is always computed between 0.3--1~keV and 1--4~keV, unless the source was not variable in the soft band to measure a lag.  In this case, we measure the lag-frequency spectrum between 2--4~keV and 4--7~keV.

The lag-energy spectra are computed with the same technique, except this time, the lag is computed at a particular frequency range between a narrow band-of-interest and a large reference band.  To obtain the best signal-to-noise, we choose the reference band to be the entire 0.3--10~keV band, and remove the band-of-interest so that the noise is not correlated.  Again, the number of fine energy bins is dictated by the quality of the data.  We aim for roughly equal logarithmic bins, though this is not always possible as many sources have a clear energy dependence to their variability.  For example, sources that are heavily absorbed will require few energy bins at low energies, but will permit finer energy binning at high energies, where there is more variability. Or, if a source spectrum is very steep, there will be many more photons at low-energies, and therefore finer energy binning can be used at lower energies. We aim to bin such that the errors on the lag are roughly equivalent in each energy bin.  

See the Methods section of the review by \citet{uttley14} for more details on the Fourier time lag analysis.

\section{Results}
\label{results}

\subsection{The lag-frequency and lag-energy spectra}

\begin{table*}
\centering
\begin{tabular}{llll|lllll}
\hline
\hd{4}{\bf Published iron~K lags} & \hd{4}{\bf iron~K lags (this work)} \\
\hline
{\bf Name } & {\bf  Reference} & {\bf  Amplitude} & {\bf log(L$_{\mathrm{bol}}$)} & {\bf  Name} & {\bf C.I.\%-A} & {\bf C.I.\%-B} & {\bf Amplitude} & {\bf log(L$_{\mathrm{bol}}$)} \\
\hline
NGC~4151        &   \citet{zoghbi12}   & $880\pm360$ &44.01	&		NGC~5506  &      $>99.999$ & $>98.5$	&	 $398\pm252$  &  44.22\\
1H~0707-495     &   \citet{kara13a}    & $47\pm16$    &44.43	&		NGC~7469  &      $>99.7$& $>99.99$	&	 $1848 \pm 1451$       &  45.10\\
IRAS~13224-3809 &   \citet{kara13b}    & $299\pm135$  &45.74	&		IC~4329A	 &      $>99.99$& $>97$	&	 $696\pm331$  &  44.92\\
MCG-5-23-16	&  \citet{zoghbi13}  & $1037\pm455$ &44.30	&		IRAS~13349 &$>80$	& $>78$	& $<370$       &  45.72\\
NGC~7314         &  \citet{zoghbi13}   & $77\pm31^{1}$    &42.98  &		IRAS~17020 &$<68^{2}$& $>97.5$	&	 $128\pm88$&	44.74\\
Ark~564          &  \citet{kara13c}    & $92\pm65$    &44.36&		IRAS~18325 &$>99.99$ 	& $>82$ &        $<153$    &      44.48\\
Mrk~335		&  \citet{kara13c}    & $193\pm98$   &45.10&		NGC~3783    &    $>99.9$ 	& $>99$ &        $172\pm62$ &    44.28\\
PG~1244+026      &  \citet{kara14a}    & $726\pm306$  &44.62	&		NGC~6860    &    $>99^{3}$& $>75$	&         $398\pm252$ &   43.71\\
Swift~J2127    &  \citet{marinucci14b}& $408\pm127$  &44.55	&		PG~1211+143 &    $>90$	& $>70$ &        $1179\pm980$  & 46.17\\
NGC~1365         &  \citet{kara15a}    & $500\pm120$  &43.99&		NGC~5548    &    $<68$   &       $>97.5$ & $311\pm109$  &  44.79\\
MS~22549-3712    &  \citet{alston15}  & $1500\pm850$ &45.09&		ESO~362-G18 &    $>99$	& $>99$&         $1562\pm606$  & 44.11\\
RE~J1034+396     &  Markevi{\v c}i{\= u}t{\.e} et al., {\em in prep.}   &  $450\pm200$ & 44.52 & & & & \\
NGC~4051         &  Alston et al., {\em in prep.}  & $90\pm30$    &43.26& & & & \\
\hline
\end{tabular}
\caption[Iron~K lags]{On the left, we show sources that have published iron~K lag results, their references, amplitudes of the iron~K lag (i.e. the lag between 3--4~keV and 5--7~keV), and bolometric luminosities.  On the right, we present new sources where there are hints of iron~K reverberation. C.I.\%-A and C.I.\%-B show the significance of the observed lag-energy spectra to two different null hypotheses (see text and Fig.~\ref{ngc5506_sim} for information on null hypothesis models). Sources must pass both tests with $>97$ per cent confidence to be considered `detections' of iron~K reverberation.  $^1$ NGC~7314 is the only source where we used a different value from the literature. Since \citet{zoghbi13}, an additional observation of NGC7314 has been taken, revealing a smaller amplitude iron~K lag.  As these are both detections of an iron~K lag, we took the average from all observations.  $^2$ For the case of IRAS~17020+4544, the continuum appears to be at 0.7--2~keV, rather than from 1--3~keV.  If instead we fit the null hypothesis from 0.7--2~keV for this source, we obtain a confidence of $>99$~per cent.  $^3$  For NGC~6860, the null hypothesis from fitting the 1--3~keV continuum yields a significance of $>8\sigma$, which is clearly too high given the data quality. If, instead, we force the null hypothesis to have a slope of 1, then the significance becomes $>99$~per cent, which is more satisfactory.}
\label{reverb_table}
\end{table*}

The time lags for all sources are shown in the Appendix, though we show one of the best examples of Ark~564 in Fig~\ref{ark564_lags}.  The time-integrated fluxed energy spectra are shown for each individual observation in the left-most column.  The time-integrated energy spectra for each individual observation are shown so the reader can see how many observations were used, can assess how bright the source is, and see the general shape of the spectrum (without doing any explicit modelling of the source).  We plot fluxed spectra (i.e. unfolded spectra with respect to a powerlaw with index 0 and normalization 1, which is equivalent to dividing by the effective area only (see \citealt{vaughan11}).  For this study we average over all observations to obtain a general picture of the time lags.  For most sources, separate observations exhibit stationarity, however there are some cases where reverberation was only found in particular observations (e.g. when the source is largely unobscured or when the source is strongly reflection dominated).  Those sources require more detailed analysis, but that is beyond the scope of this work.

The next panel ({\em left-middle}) of Fig.~\ref{ark564_lags} shows the lag-frequency spectrum between 0.3--1~keV and 1--4~keV, which are typically the bands dominated by the soft excess and continuum, respectively.  These two bands have been very successful in revealing high-frequency reverberation (e.g. \citealt{fabian09}, \citealt{cackett13} and \citealt{demarco13}).  While these may not be the ideal bands for some sources (as the energy of the continuum and soft excess are not necessarily the same for all sources), we chose to use the same bands for all sources in order to have consistency throughout the sample. There are a few sources (e.g. MCG-5-23-16, NGC~7314) where it is not possible to measure a significant lag between the soft and the hard bands because the soft band is not highly variable.  These are usually cases where there is strong neutral absorption along our line of sight.  In these cases, we measured the lag between the more variable hard bands (between 2--4~keV and 4--7~keV).  The sources where this was required are indicated by two asterisks (**) in the figures.  The analysis in the 2--4~keV and 4--7~keV bands reveals the frequency at which there is a time lag, however, it does not easily distinguish the low-frequency continuum lag from the reverberation lag at high frequencies because in this projection, both continuum and reverberation lags will be seen as positive, hard lags.

The panels on the right show the lag-energy spectra. The frequency ranges for the lag-energy spectra were chosen based on the lag-frequency spectra.  This minimizes the bias associated with hand-picking the frequency at which one finds an iron~K lag.  For some sources, the frequency range of the low-frequency lag is clearly distinguished from the high-frequency reverberation lag (e.g. 1H~0707-495, Ark~564).  In other sources it was only possible to look at the lags in one frequency band, either because the statistics were poor and we needed to average over a wide frequency range, or because there was only evidence for one process dominating the lags (i.e. the lag-energy spectra did not change shape for the entire available frequency range).   The blue and red arrows in the lag-frequency spectra show the frequency at which the softest and hardest bands of the lag-energy spectra become dominated by Poisson noise.  Often the frequency at which the hardest band became dominated by Poisson noise (as indicated by the red arrow) dictated the upper bound of the high-frequency lag-energy spectrum.  This is a conservative approach, as the cross spectrum picks out correlated variability between the reference band and the bin-of-interest, even if the power of the bin-of-interest is dominated by Poisson noise.  
These prescriptions allowed for us to look at the lags in a large number of sources. It is not necessarily the ideal frequency or energy binning for each individual source, but it does allow for some consistency between sources.

\subsection{Characterizing the lags}

To guide the reader, the low- and high-frequency lag-energy spectra for each source are marked with colored markers to indicate what rough catagory they appear to fit into.  Magenta hexagons indicate sources with iron~K reverberation that have been previously published. Cyan squares indicate sources with new evidence for iron~K reverberation found from this study at varying levels of significance.  Blue circles indicate the low-frequency hard lag, and red diamonds indicate low-frequency soft lags that are not obviously associated with reverberation (i.e. have no indication of iron~K reverberation).  A few lag-energy spectra do not fit into any category, for example, the high-frequency lag-energy spectrum of Mrk~766 and the low- and high-frequency lag-energy spectra of ESO~113-G010, NGC~3516, and NGC~4395.

There are currently 13 sources with published iron~K reverberation lags (see Table~\ref{reverb_table} for more detail and references).  These sources all show a clear drop in the lags above the iron line at $\sim 7.5$~keV.  It is important to emphasize that some of the published results on iron~K reverberation were from specific observations, and not from all available observations, as is shown in this large sample.  For example, NGC~1365 and NGC~4151 are sources that show clear neutral absorption.  Iron~K reverberation was found only in observations taken when the source was largely unobscured.  For IRAS~13224-3809 and NGC~4051, iron~K reverberation was only found when these sources were in a low-flux state.  Therefore, the lags from this large sample do not necessarily present the best reverberation results, rather they show the average time lags from all available observations.  Several of the sources in this sample require a more detailed analysis, but this is beyond the scope of this project. 

Eleven other sources in our sample show hints of reverberation, in that the lag-energy spectrum peaks at around 6--7~keV, the line centroid of the iron~K emission line.  For these sources, we assess the significance of this iron~K feature based on a simple $\chi^2$ goodness of fit comparison to a simple log-linear model null hypothesis.  For the reverberation signature, we want to know: does this lag-energy spectrum differ significantly from either zero lag at all energies or from a featureless change in the lag with energy.  In particular, with reverberation, we expect the reflection feature to show some deviation from the lags in the continuum band.  Therefore, we fit a simple log-linear model to the 1--3~keV continuum dominated band (i.e., $y=a+b~\mathrm{log}(x)$).  We extrapolate this simple model to the 3--10~keV, and calculate the $\chi^2$ in the 3--10~keV band.  The 1--3~keV band is a sensible choice for the null hypothesis model because (1) it is the band dominated by the direct powerlaw emission and (2) it is the band at the peak of the {\em XMM-Newton} effective area, and therefore the error bars on the lag are relatively small in this band.  For sources that have 1 or 2 energy bins in the 1--3~keV band (NGC~5506, NGC~3783, IRAS~13349+2438), the null hypothesis assumed is a flat line with a slope of 0, indicating zero lag in the continuum band. The confidence levels determined from a $\chi^2$ distribution are shown in column C.I.\%-A in Table~\ref{reverb_table}.

\begin{figure}
\includegraphics[width=\columnwidth]{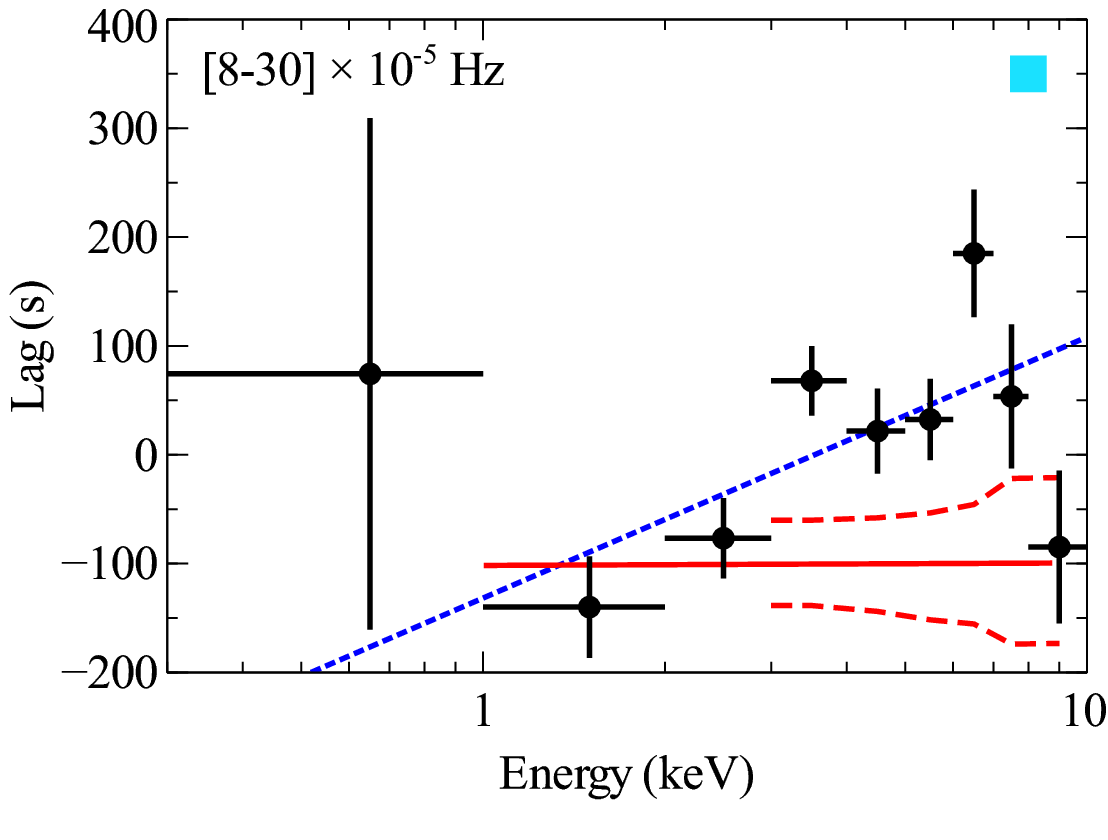}\\
\caption{An example of the technique for measuring the significance of the iron~K lag for the case of partially obscured Seyfert 1.9 galaxy, NGC~5506.  Here we show the high-frequency lag-energy spectrum of the source (see Appendix for the low-frequency lag-energy spectrum, which shows a featureless increase with energy above 2~keV).  The red solid line shows the null hypothesis fit in the 1--3~keV band and extrapolated to the 3--10~keV band. The dashed red contours show the $1 \sigma$ spread in lags from Monte Carlo simulations (see text for details).  The simulations confirm that the data differ from the null hypothesis model at $>99.999$ per cent confidence.  We also provide a second statistical test in which the null hypothesis is a log-linear fit to the entire 0.3--10~keV band (blue dashed line). We require that a source pass both tests at $>97$\% confidence to be considered a `detection' of iron~K reverberation.} 
\label{ngc5506_sim}
\end{figure}

We show an example of this statistical test for the case of NGC~5506 in Fig.~\ref{ngc5506_sim}.  Here we show the high-frequency lag-energy spectrum, which shows evidence for an iron~K lag.  We fit the 1--3~keV continuum dominated band with a flat line (because the 1--3~keV band consists of only two bins), and extrapolate it up to 3--10~keV iron line band. This null hypothesis model is shown as the red solid line.  Then the $\chi^2$ is calculated between the null hypothesis and the bins in the 3--10~keV band.  In this example, we have also confirmed the $\chi^2$ approach with Monte Carlo simulations. The red dashed lines show the $1 \sigma$ spread in lag from 1000 Monte Carlo light curve pairs in each energy band.  We use the method described in \citet{timmer95} to produce Monte Carlo light curves based on the observed PSD in each energy band.  The simulated light curves are scaled to the mean and variance of the observed light curves.  Our simulations show that the observed lags differ from the null hypothesis at $>99.999$ per cent confidence (as the $\chi^2$ analysis also showed). The simulations also confirm that the $1 \sigma$ error bars that we calculated using the method in \citet{nowak96} are the same as those determined from a Monte Carlo analysis (see also \citealt{uttley14} and \citealt{demarco13}).  

As another statistical check, we tested how much the observed lag-energy spectrum differs from a simple log-linear model fit to the entire 0.3--10 keV band.  An example of this test for the case of NGC~5506 is shown as the blue dotted line in Fig.~\ref{ngc5506_sim}, and the confidence levels determined from the fit are shown in C.I.\%-B in Table~\ref{reverb_table}.  We consider sources to have `detections' of iron~K reverberation if they pass both statistical tests at $>97\%$ confidence.  There are 7 sources that meet this criterion.

While these statistical tests and the Monte Carlo simulations confirm our error bar size and the presence of some structure in the lag-energy spectra, there are some difficulties in measuring the significance of specifically the iron~K in the lag-energy spectra.  The biggest source of systematic uncertainty is in our choice of the null hypothesis.   We do not yet have a model to describe the high frequency lags from the continuum alone, and therefore it is difficult to find a model to compare against the iron~K feature.  A small source of error also arises from our choice of the $\chi^2$ statistic, which assumes that all data points are independent.  The lags are measured to a common broad reference band with the bin-of-interest removed, and therefore technically, there is some common denominator between all points. However, the error on each bin-of-interest is dominated by the noise in this narrow bin and not by the common broad reference band. Therefore the error associated with correlated data points is minimal, especially compared to the larger systematic uncertainty in the null hypothesis.  Overall, we present the reader with the probability against these null hypotheses as an aid, but we emphasize the importance of considering the lags in the larger context.  The fact that these iron~K lag features peak at 6--7~keV, are commonly accompanied by a soft excess lag of a similar amplitude, and the fact that the lag-energy spectra at lower frequencies show a featureless increase with energy, all support the detection of iron~K reverberation.

In Fig.~\ref{fvar_reverb} we fill in the rms-counts plane with sources that show iron~K reverberation (either from previously published works, in magenta hexagons, or from this work, in cyan squares).  One can see that most of the sources in the top-right and portion of the plot (i.e. statistically the `best' sources) have been studied in detail in previous work. Through this large sample, we are filling more sources reaching to the lower-left portion of the plot.  20/43 sources show reverberation (or about half), but we emphasize that in some previous work, reverberation was only found in particular flux states (e.g. \citealt{zoghbi12,alston13,kara13b}), and therefore it is possible that other sources have reverberation that was not revealed through our more general study. 

Some sources in the top-right portion of the plot do not show clear signs of reverberation, despite their statistical favorability.  We have labeled a few of these sources, including most notably MCG-6-30-15 and Mrk~766.  We performed a detailed study of the lags in MCG-6-30-15 \citep{kara14b}, and found that while the source showed a low-frequency hard lag that increased steadily with energy, there was only very weak evidence for an iron~K lag.  A comparison of the time-integrated energy spectrum and the covariance spectrum (i.e. the spectrum of the correlated variability at a particular frequency, a.k.a. the spectrum of the emission that contributes to the lag-energy spectrum) showed that while a broad iron~K line was present in the spectrum, it was not varying in a correlated way with the continuum emission.  If the reflection spectrum is not varying with the continuum, no lag will be detected. Detailed follow-up using other spectral-timing measures is required to understand if this effect is present in other sources, as well.

Most of the sources show hard lags that increase steadily with energy up to 10~keV.  29/43 ($\sim 70$~per cent) sources visually show low-frequency hard lags, where there is a log-linear increase in the lags from 2--10~keV. We indicate these lag-energy spectra with blue circle markers in the Appendix.  The sources with low-frequency hard lags are shown in blue in Fig.~\ref{fvar_hard} on the rms-mass relation.

Of the remaining 14 without clear low-frequency hard lags, 8 sources show lag-energy spectra that suggest that an iron~K lag is present at low frequencies (MCG-5-23-16, NGC~7314, PG~1244+026, IC~4329A, PG~1211+143, SWIFT~J2127.4+5654, NGC~4151 and ESO~362-G18; shown as magenta hexagons in Fig.~\ref{fvar_hard}). Most of these sources have a higher black hole mass, meaning that the reverberation lag occurs at lower frequencies.  The remaining six sources (shown as unfilled circles in Fig.~\ref{fvar_hard}) do not show low-frequency hard lags and do not show any evidence for reverberation in the form of a low-frequency iron~K lag. These include NGC~1365, NGC~3516, NGC~4395, ESO~113-G010, Mrk~586, and Mrk~841. For the case of NGC~1365, where we see high-frequency reverberation in addition to a low-frequency soft lag, in \citet{kara15a} we interpreted the low-frequency lag as due to variable neutral absorption.  A more detailed look at these six sources is required to ascertain whether absorption is affecting their low-frequency lags or if these sources harbour high mass black holes ($\sim 10^8 M_{\odot}$ or greater), but the quality of the data is not high enough to detect reverberation.  Several of these sources show complex and variable absorption, and do not have a particularly large black hole mass (e.g. NGC~3516 and NGC~4395).

Curiously, four sources show the unexplained behaviour of the featureless hard lag that disappears at the lowest frequencies.  This was first seen in MCG-6-30-15 \citep{kara14b}, and we see similar behaviour in NGC~3227, Mrk~205 and Mrk~1040.    

Finally, 21 of 29 with low-frequency hard lags show a different shape at high frequencies. While it is not always clear if this shape indicates reverberation, it is clear that there is a change from low to high frequencies.

In summary:
\begin{enumerate}
\item 13 sources have published iron~K reverberation results. Now, including the 7 new sources with iron~K reverberation with $>97$~per cent confidence, we find reverberation in 20/43 sources or $\sim 50$~per cent of sources.
\item The low-frequency hard lag (commonly interpreted as propagation lags) is common in 29/35 sources (excluding sources that show reverberation down to the lowest frequencies probed).  This is $83$ per cent of sources.
\item There are 6 sources in the sample that have low-frequency lags that do not resemble the low-frequency propagation or reverberation.  Further work is required to test if these lags are associated with variable absorption on timescale of tens of kiloseconds (Silva et al., {\em in prep.}).  
\end{enumerate}

\section{Discussion}
\label{discuss}

\subsection{Reverberation correlations with mass and mass accretion rate}

The time lag between the continuum and the reflected iron~K emission is caused by the light travel time between the corona and the ionized accretion disc.  Therefore, if all systems roughly have a similar coronal geometry, we expect that the amplitude of the lag will scale with black hole mass.  This reverberation lag-mass scaling relationship has been shown for the soft lags \citep{demarco13}, and for a few sources with iron~K lags as well \citep{kara13c}.  The scaling relation also extends to reverberation in black hole binaries \citep{demarco15}.

\begin{figure}
\includegraphics[width=\columnwidth]{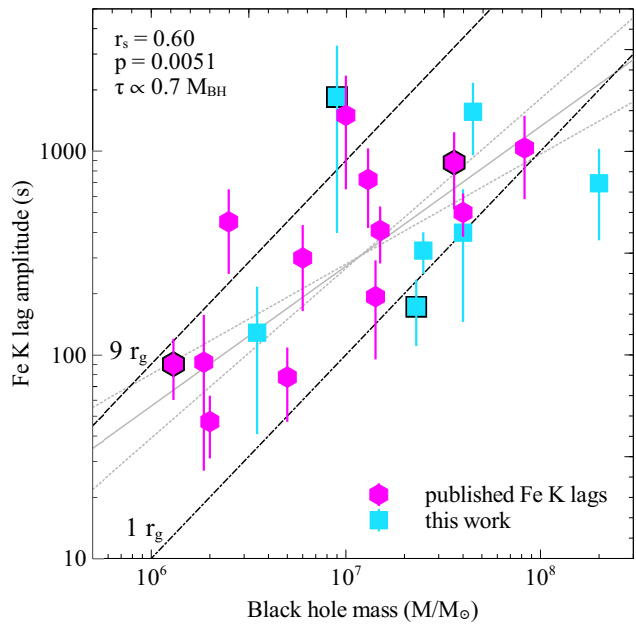}\\
\caption{Iron~K lag amplitude vs. mass for sources with previously detected iron~K lags (magenta) and ones detected in this work (cyan).  See text for how amplitude of the lag was measured for all sources.  The black diagonal dot-dashed and dashed lines show the time delay at 1 and 9~$r_{\mathrm{g}}$. The grey solid line indicates the best fit linear model (on a log-log plot), and the grey dashed lines indicate the 1-$\sigma$ errors on the free parameters. Sources with Optical reverberation mass estimates are highlighted with black outlines. We find a Spearman Rank Order coefficient, $r_{\mathrm{s}} = 0.60$ at a probability of 0.0051.  The correlation coefficient for just previously published iron~K lags (magenta hexagons) is 0.69 with a p-value of 0.0098.}
\label{updated_lag_mass}
\end{figure}

In Fig.~\ref{updated_lag_mass} we show the iron~K lag-mass relation for the 13 sources with published iron~K reverberation measurements in magenta hexagons.  The lag has been measured between the 3--4~keV band and the 5--7~keV band.  We use the results from the literature because the results on these individual sources are likely to be the most complete analysis performed.  We use mass estimates from the literature, as discussed in Section~\ref{sample_section}. Only 4 of the 20 sources in our sample have optical reverberation mass estimates. For clarity, we do not plot the error on the mass, though the errors are included when fitting the model. We use an Orthogonal Distance Regression fitting procedure to account for the error in both x and y variables.  For these sources with previously published iron~K results, the Spearman rank order correlation coefficient is 0.69 with a p-value of 0.0098.

This study has allowed us to detect iron~K reverberation in a larger sample of sources. The cyan squares in Fig.~\ref{updated_lag_mass} show the new reverberation measurements (greater than 97~per cent confidence level) found through this analysis.  Some of these sources have lower signal-to-noise reverberation signatures. For these sources, it is not always possible to measure the lag between the 3--4~keV bin and the 5--7~keV bin, due to the large energy bins required for the lower quality data.  Instead, we calculated the difference between the bins with the smallest and largest lags in the 2--7~keV band. With this study, we are filling out the higher-mass end of the relation, where reverberation is typically more difficult to measure.  For all 20 sources with detected iron~K lags, we find a Spearman Rank Order Correlation coefficient of 0.60 with a p-value of 0.0051. The grey solid line shows the best fit linear model (in log-log space) to all 20 sources, and the dashed lines show the $1 \sigma$ errors on slope and y-intercept. The four sources with Optical reverberation mass estimates do not show a stronger correlation than other sources, which could indicate that the spread we observe is due to intrinsic difference in the source/accretion disc geometry, rather than errors in the mass estimates.

\begin{figure}
\includegraphics[width=\columnwidth]{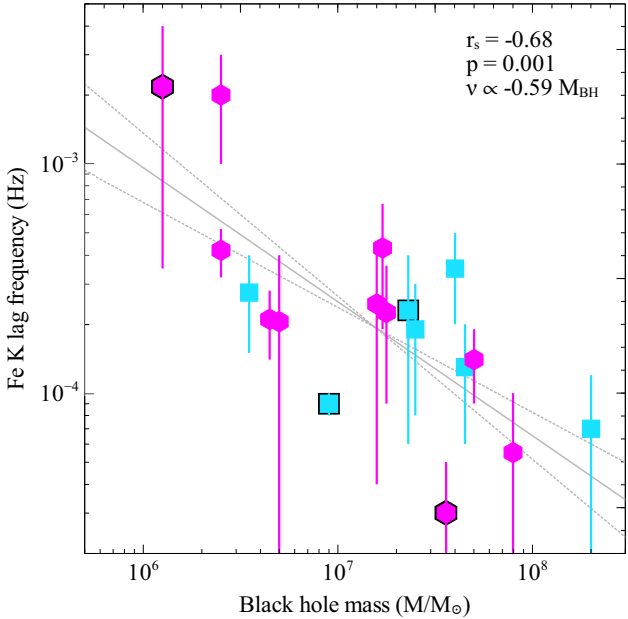}\\
\caption{The temporal frequency at which the iron~K lag is measured vs. mass for sources with previously detected iron~K lags (magenta) and ones detected in this work (cyan).  The grey solid line indicates the best fit linear model (on a log-log plot), and the grey dashed lines indicate the 1-$\sigma$ errors on the free parameters. Sources with Optical reverberation mass estimates are highlighted with black outlines. We find a Spearman Rank Order coefficient, $r_{\mathrm{s}} = -0.68$ at a probability of 0.001. The correlation coefficient for just previously published iron~K lags is -0.74 with a p-value of 0.0037.}
\label{freq_mass}
\end{figure}

For completeness, the same analysis is shown in Fig.~\ref{freq_mass} for the temporal frequency at which the iron~K lag is found, and the black hole mass.  Similar results have been found for the frequency of the soft lag in \citet{demarco13}.  As discussed earlier, the frequency range of the iron~K lag is determined by the frequency at which there is a soft lag. The y-axis error bars in Fig.~\ref{freq_mass} shows the frequency range, and the y-axis value is the midpoint of that frequency range.  The correlation coefficient is stronger for the frequency-mass relation than for the lag-mass relation, which suggests that it is better to use the frequency-mass relation when determining the black hole mass of an object using X-ray reverberation mapping. 

While the reverberation lag is caused by the light travel distance between the corona and the disc, we cannot simply convert the amplitude of the lag that we measure into a light-travel distance.  This is mostly because of dilution effects.  Both the 3--4~keV band and the 5--7~keV band are composed of continuum and reflected emission.  As a simple example: if both bands were composed of 50~per cent direct continuum and 50~per cent reflected emission, then we would measure zero lag between the bands, regardless of the true light travel time delay between the corona and the disc.  However, in the reverberation cases we see here, the 3--4~keV band has a higher fraction of continuum emission than the 5--7~keV band, which is why we measure a lag between the bands.  Dilution will always cause the measured lag to be less than the true light travel time delay between the corona and the disc.  Many effects, such as coronal geometry and ionization parameter can effect the reflection fraction (and therefore the amount of dilution), so it is important to account for these effects when converting the lag into a distance \citep{chainakun15}. While dilution is important, we note that independent spectral modelling of some of these sources indicates that this effect is usually not greater than a factor of four (e.g. \citealt{kara15a}). For this simple lag-mass relation, we have not accounted for dilution effects because we wanted to show the model-independent results.  The diagonal lines show the mass-invariant distances in gravitational radii for light travel distances from a point source above the disc irradiating a face-on accretion disc. All the sources lie within 10~$r_{\mathrm{g}}$, but we emphasize again that we have not corrected for dilution, and so these values do not reflect the true light travel distance.  Nonetheless, what is clear, is that the reverberation time delay is short for all sources, suggesting a compact corona close to the central black hole.  Independent results from microlensing of distant quasars also show that the X-ray emitting corona is compact (within $10~r_{\mathrm{g}}$) compared to the larger Optical emitting region (e.g. \citealt{chartas12}).

There is clearly a large spread in this relation that we need to understand.  Some of the spread will be because we have not modelled the dilution in these sources, as discussed in the previous paragraph.  Some of the spread is likely physical.  For example, some sources may have very compact coronae at a height of $2 r_{\mathrm{g}}$, while others may be more extended and therefore the reverberation lag will be longer.  On the other hand, some source of the scatter is due to observational biases.  The amplitude of the lag depends on the frequency at which the lag was measured.  The frequency range can be skewed because it is dependent on the length and time resolution of our data.  Sources with higher mass have longer and slower reverberation signatures that may be too long for our observations to detect (this is likely what is happening in the case of IC~4329A, which shows a very short reverberation lag for its large $2\times10^{8} M/M_{\odot}$ black hole).  The way to disentangle these physical and observational forms of scatter is to model the reverberation (e.g. \citealt{cackett14}, \citealt{emm14}, and \citealt{chainakun15}).


\begin{figure*}
\includegraphics[width=0.66\textwidth]{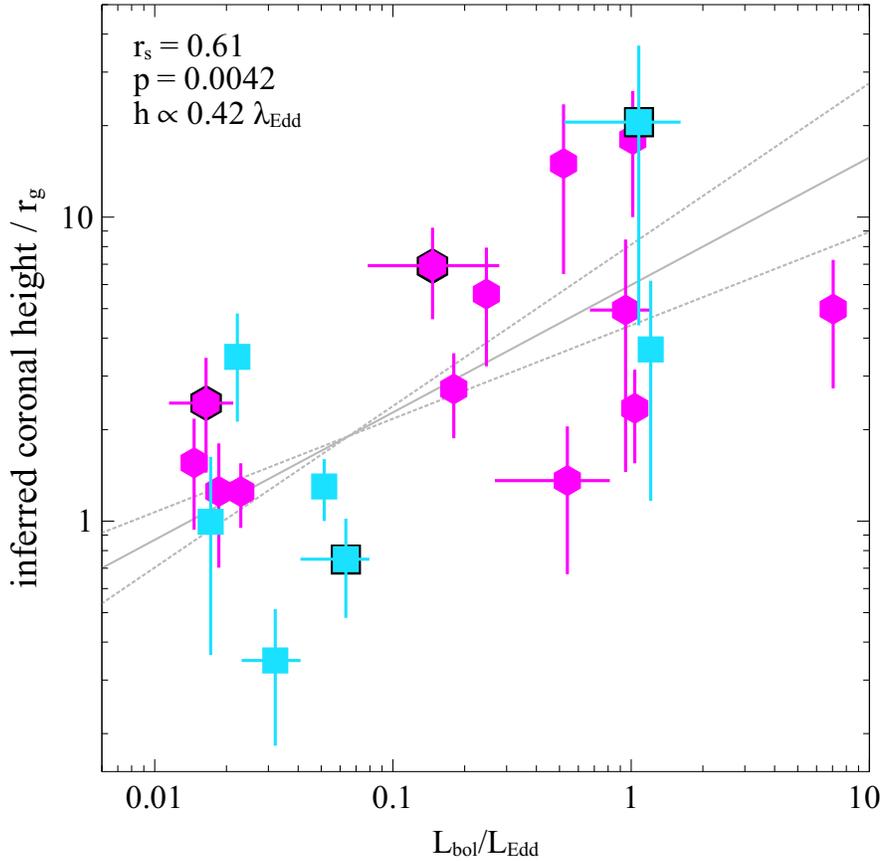}\\
\caption{The inferred coronal height in units of $r_{\mathrm{g}}$ vs. the Eddington ratio for the sources with iron~K reverberation detections in our sample. Th
e coronal height is simply calculated assuming that the iron~K lag amplitude (without dilution correction) is the light crossing time delay between a point sour
ce corona and a face-on disc.  Again, magenta hexagons refer to sources with previously detected iron~K lags, while cyan squares refer to new reverberation sour
ces from our sample. Sources with Optical reverberation mass estimates are highlighted with black outlines. The errors on both axes do not include the error ass
ociated with mass (as both axes are computed by dividing by the mass). The error in Eddington ratio is simply the range of values found in the literature. Sourc
es with only one estimate are shown without error bars on this plot, but we included an error of 0.2 dex when computing the best-fit linear model (grey lines).
 The Spearman Rank Order Correlation Coefficient is 0.61 with a p-value of 0.0042. For previously published iron~K lags (magenta) the correlation coefficient is
 lower: 0.43 with a p-value of 0.14.  We note that we find no significant correlation between $r_{\mathrm{g}}$ and bolometric luminosity (see text for details).
}
\label{rg_eddington}
\end{figure*}

With this growing sample of sources, we can begin to examine other possible reasons for the potential spread in coronal source heights.  In Fig.~\ref{rg_eddington}, we show the inferred coronal height (in units of gravitational radius) vs. Eddington ratio for the 20 reverberation sources.  The coronal height is calculated assuming that the iron~K lag amplitude is the light travel time between a point source, on-axis corona and a face-on disc. It does not account for dilution.  Again we caution the reader against taking the value in $r_{\mathrm{g}}$ as the actual source height, but it it is a convenient way of thinking about the reverberation lag without the dependence on mass.  We look for a possible correlation with Eddington ratio, which is also independent of mass.  The Eddington ratio was calculated using the bolometric luminosities calculated from the SEDs in \citet{vasudevan07,vasudevan09,vasudevan10}, and \citet{wang04}. Where there were multiple estimates of $L_{\mathrm{bol}}$, we took the mean.  There were no estimates of $L_{\mathrm{bol}}$ for Swift~J2127.4+5654, MCG-5-23-16, IRAS~18325-5926 or ESO~362-G18, and so for those, we relied on estimates from \citet{miniutti09}, \citet{beckmann08}, Iwasawa et al., {\em submitted}, and \citet{agis14} respectively.  The y-error bars show the error associated with the lag amplitude and the x-error bars show the spread in values for the bolometric luminosity. We do not include the error associated with mass in either x or y variables.  For sources with only one estimation of the bolometric luminosity, we assume an error of 0.2 dex when fitting the best fit linear model in log-log space (shown as the grey solid line).  There is large scatter in this plot, though there are hints that the higher mass accretion rate sources have longer reverberation time delays. We find a Spearman Rank Order Correlation Coefficient of 0.61 with a p-value of 0.0042.

\begin{figure*}
\includegraphics[width=\textwidth]{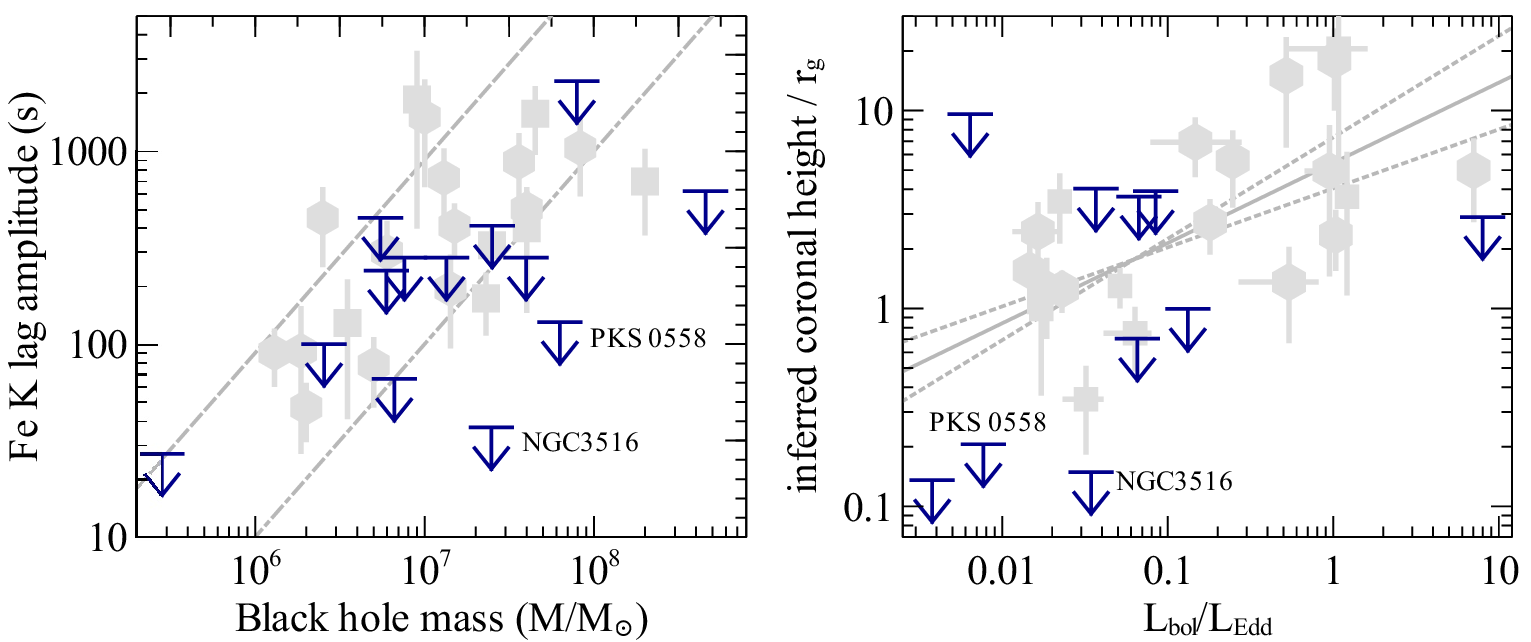}\\
\caption{{\em (left:)} The iron~K lag amplitude vs. black hole mass relation as in Fig.~\ref{updated_lag_mass} (in gray points), now including upper limits from sources with non-detections (dark blue arrows). See text for details on how upper limit on lag is measured.  The dot-dash and dashed lines correspond to the light-crossing time at 1 and $9~r_{\mathrm{g}}$, respectively.  Two sources seem to diverge from the correlation most, NGC~3516 and PKS~0558-504.  {\em (right:)} Same as Fig.~\ref{rg_eddington}, now including upper limits in dark blue.  The relatvely small lag of NGC~3516 and PKS~0558-504 may be explained by their low Eddington ratios.} 
\label{upperlimits}
\end{figure*}

As both Eddington ratio and coronal height are computed by dividing by the mass, a large error on the mass would shift the points in both x and y variables, and could therefore produce a false correlation.  However, our mass estimates are known to within a factor of a few, and therefore this is unlikely.  We confirm this by testing the correlation between $r_{\mathrm{g}}$ and bolometric luminosity. We find no correlation.  The Spearman Rank Order Correlation Coefficient is 0.33 at a probability of 0.16.  This gives us confidence in the correlation shown in Fig.~\ref{rg_eddington}.

Lastly, we ask: are sources with non-detections of reverberation consistent with these correlations? This can be a challenging question to answer as there are different processes occuring at different frequencies, and therefore we must be careful that we are looking at the frequency range that contains reverberation (rather than, for instance, the hard continuum lag).  To account for this, we look for sources that have low-frequency hard lags that go to zero at the highest frequencies.  Previous results (like those in Fig.~\ref{ark564_lags}) show that the reverberation lag occurs at higher frequencies than the hard lag, and because of this, we postulate that in these sources with non-detections, reverberation would be at these high frequencies if the data quality were high enough.  There are 13 sources that show clear hard lags at low frequencies, but do not show significant reverberation at high frequencies.  Many of these sources show $<1\sigma$ soft lags at high frequencies. Using the frequency at which we see this tentative soft lag, we calculate the lag between the 3--4~keV band and the 5--7~keV band (as we did for the correlation in Fig.~\ref{updated_lag_mass}).  The upper limit then comes from the 1-$\sigma$ error on the lag measured between 3--4~keV and 5--7~keV at the frequency where a tentative soft lag is found. MCG-6-30-15 is the only source that shows a low-frequency hard lag and a soft lag at high frequencies, but also shows a significantly soft lag between the 3--4~keV an 5--7~keV bands (thus inhibiting a measurement of the upper limit of an iron~K lag).  

We show the iron~K lag amplitude vs. black hole mass (Fig.~\ref{updated_lag_mass}) and the coronal source height vs. Eddington ratio (Fig.~\ref{rg_eddington}) with the upper limits from 13 sources in Fig.~\ref{upperlimits}.  The upper limits are largely consistent with the correlations, perhaps with the exception of NGC~3516 and PKS~0558-504.  This may be suggesting that the high-frequency variability in these sources is not associated with reverberation.  Alternatively, the lags in these two sources may be relatively `small' because they have low Eddington ratios. This is perhaps suggested by the right panel of Fig.~\ref{upperlimits}.  Both NGC~3516 and PKS~0558-504 are consistent with the trend of small lag for low Eddington ratio objects.

It is possible that these 13 sources with non-detections would contain iron~K lags if we had high enough data quality, or it is possible that these sources have different corona/disc structures that inhibit the measurement of a clear iron~K lag.  Understanding these correlations in greater detail will be important for understanding the structure of the corona and its connection with the mass accretion rate and disc geometry.


\subsection{Low-frequency lags}

Most of the low-frequency lags in our sample can be understood within the framework of either the propagating fluctuations model (25/43 or $\sim 60$~per cent of sources) or in terms of reverberation that goes down to the lowest frequencies we can probe (8/43 or $\sim 19$~per cent of sources).  The remaining 10 sources are: NGC~1365, NGC~3516, NGC~4395, NGC~3227, MCG-6-30-15, Mrk~841, Mrk~586, Mrk~205, Mrk~1040 and ESO~113-G010. Four of these sources, MCG-6-30-15, NGC~3227, Mrk~205 and Mrk~1040, show evidence for a hard lag that increases with increasing energy (i.e. they look like other low-frequency hard lags), but the hard lag does not extend down to the lowest frequencies.  For the remaining six sources, there is no evidence for a hard lag that increases with energy.  

Low-frequency soft lags have been shown previously in three cases of bright, variable, well-observed sources: NGC4051 \citep{alston13}, MCG-6-30-15 \citep{kara14b} and NGC~1365 \citep{kara15a}.  The low-frequency soft lag appears under different circumstances for these three sources, and it is possible that some of the others in this sample behave in the same way.  

NGC~4051 is highly variable and has a low-mass black hole ($\sim 2 \times 10^6 M_{\odot}$).  \citet{alston13} found that there was a flux dependence to the time lags in this source.  By looking at short segments (10~ks) where the source was in a high flux state, they found that the source behaved in the `normal' way (i.e. low-frequency hard lag, high-frequency reverberation).  However, the low-frequency hard lag was observed to disappear, and become a low-frequency soft lag as the source flux decreased.  The short, low-flux segments were taken from a range of different epochs, and all low-flux segments showed systematically the same behaviour of a low-frequency soft lag.  
 
NGC~1365 was recently observed by {\em XMM-Newton} for several orbits \citep{risaliti13}.  High-frequency iron~K reverberation was found during the most unobscured observation.  In this observation, there was a low-frequency soft lag.  The other observations, however, did not show the same low-frequency behaviour.  Two observations showed a low-frequency hard lag, and another had evidence for another low-frequency soft lag.  The low-frequency lags could be understood in terms of changes in the neutral absorber.  Time-resolved spectroscopy from \citet{walton14} showed that the column density of the absorber was decreasing throughout the most unobscured observation (as if the source was becoming less obscured with time).  This was found to explain the low-frequency soft lags because the hard photons could penetrate the absorber before the soft photons could (see \citealt{kara15a} for more details on the model).  A variable neutral absorber then gives a natural reason for why the low-frequency lag was different in the other observations.  This low-frequency lag differs from the one in NGC~4051 because it was not a systematic behaviour that occurred over many segments.  

Lastly, MCG-6-30-15 is a source that has been observed by {\em XMM-Newton} several times, and all observations show roughly the same mean count rate.  The low-frequency lags are also stable over all observations.  MCG-6-30-15 shows a low-frequency hard lag, similar to most sources, but at around $1.5 \times 10^{-4}$~Hz, it turns over sharply to a low-frequency soft lag.  This is different from the low-frequency soft lags in NGC~1365 and NGC~4051 in that it is present in all observations and at all flux regimes.  Also, unlike NGC~4051 where the low-frequency hard lag disappears, in MCG-6-30-15, the hard lag is always present. 

The fact that the behaviour of the low-frequency soft lag in these sources occurs under different circumstances suggests that they are not due to the same phenomenon.  
 In our sample, there are sources that appear to be more like NGC~1365, in that they show no hard lag, and ones that are more similar to MCG-6-30-15, where a hard lag is present, but turns over at lower frequencies.  Because in our systematic analysis we have averaged over all observations and all flux regimes, we cannot assess whether sources behave like NGC~4051, where the hard lag disappears at low fluxes. Also, of the 8 sources that show this interesting low-frequency soft lag (not including MCG-6-30-15 and NGC~1365), only one source, NGC~3516, has more than one variable, long observation.  This makes it impossible to test if the low-frequency soft lags are steady or change with time.  More orbit-long observations are required to understand the lowest-frequency time lags, and to understand how these lags behave over time. 

Interestingly, 4 of the subsample of 10 sources with low-frequency soft lag are known to have complex and variable neutral absorption.  These include: NGC~1365 \citep{risaliti05a,risaliti13}, NGC~4395 \citep{king13}, NGC~3227 \citep{markowitz09} and NGC~3516 \citep{turner11}.  \citet{parker15} also found that variable neutral absorption could explain much of the low-frequency variability below 2~keV in the Principle Component Analysis of NGC~1365, NGC~4395 and NGC~3227.   

\section{Conclusions}

We have presented the results from a systematic analysis of the X-ray time lags in Seyfert galaxies observed with the {\em XMM-Newton} observatory. Reverberation is found in $\sim$50~per cent of sources, and the featureless hard lag is found in $\sim 85$~per cent of sources.  We confirm the iron~K lag amplitude--black hole mass relation, and show that there are hints of a correlation between source height or extent and mass accretion rate.  X-ray spectral timing analysis is a powerful tool for probing the inner regions around supermassive black holes, and is revealing relativistic reverberation in a significant fraction of variable Seyfert galaxies. 

\section*{Acknowledgements}

Thank you to the annonymous referee for helpful comments that improved this work. EK thanks Ari Laor for interesting discussions on this work and acknowledges support from the International Space Science Institute.  This work is based on observations obtained with {\em XMM-Newton}, an ESA science mission with instruments and contributions directly funded by ESA Member States and NASA.  EK thanks the Gates Cambridge Scholarship and the Hubble Fellowship Program. Support for Program number HST-HF2-51360.001-A was provided by NASA through a Hubble Fellowship grant from the Space Telescope Science Institute, which is operated by the Association of Universities for Research in Astronomy, Incorporated, under NASA contract NAS5-26555. EK, WNA, ACF acknowledge support from the European Union Seventh Framework Programme (FP7/2007-2013) under grant agreement n.312789, StrongGravity. CSR acknowledges support from NASA under grant NNX14AF86G. EMC gratefully acknowledges support from the National Science Foundation through CAREER award number 1351222.

\bibliography{ms_arxiv}   
\bibliographystyle{mnras.bst}      

\section*{Appendix}

The following plots show the individual reasults for all 43 sources in our sample. See main text for details on the analysis.

\begin{figure*}
\includegraphics[width=\textwidth]{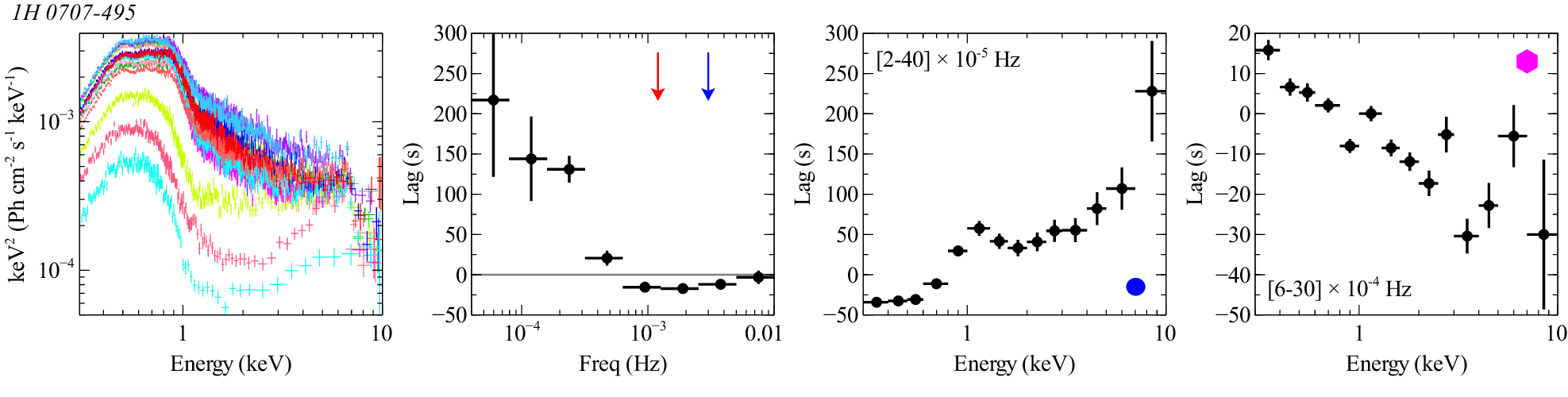}
\includegraphics[width=\textwidth]{ARK564.ps}
\includegraphics[width=\textwidth]{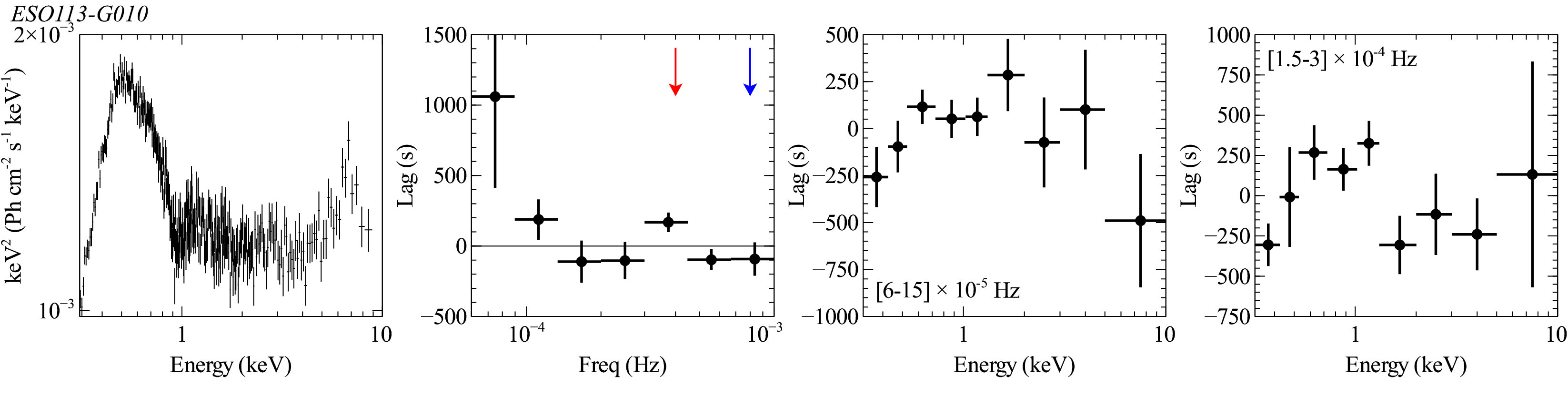}
\includegraphics[width=\textwidth]{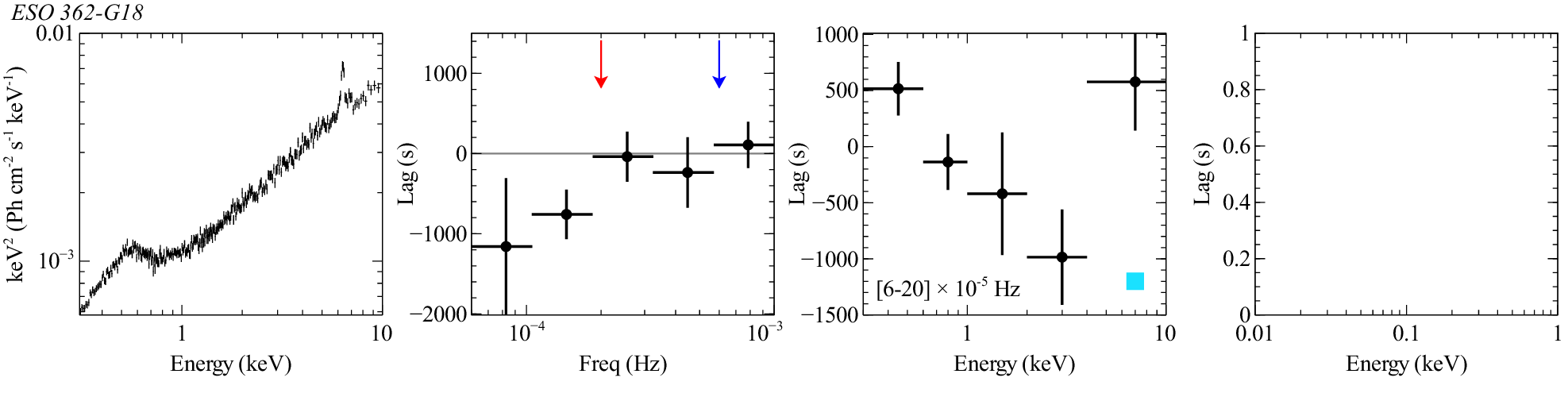}
\includegraphics[width=\textwidth]{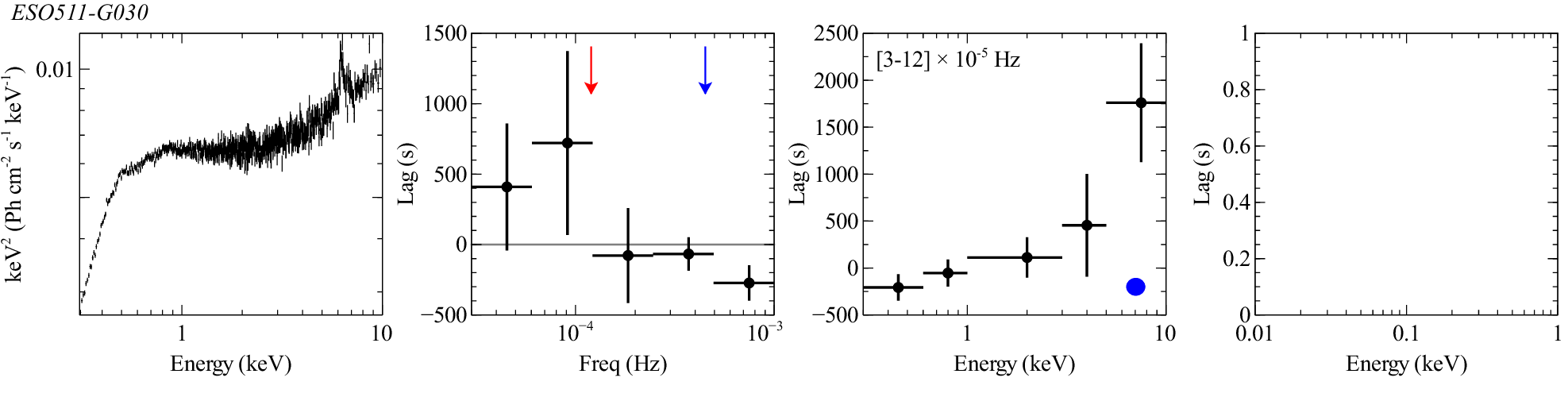}
\end{figure*}

\begin{figure*}
\includegraphics[width=\textwidth]{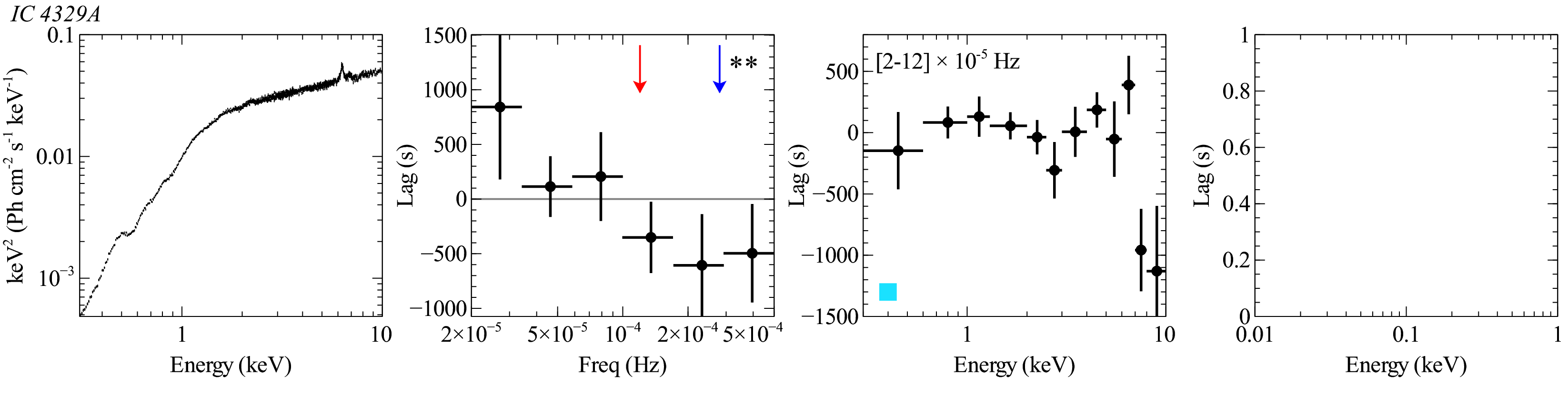}
\includegraphics[width=\textwidth]{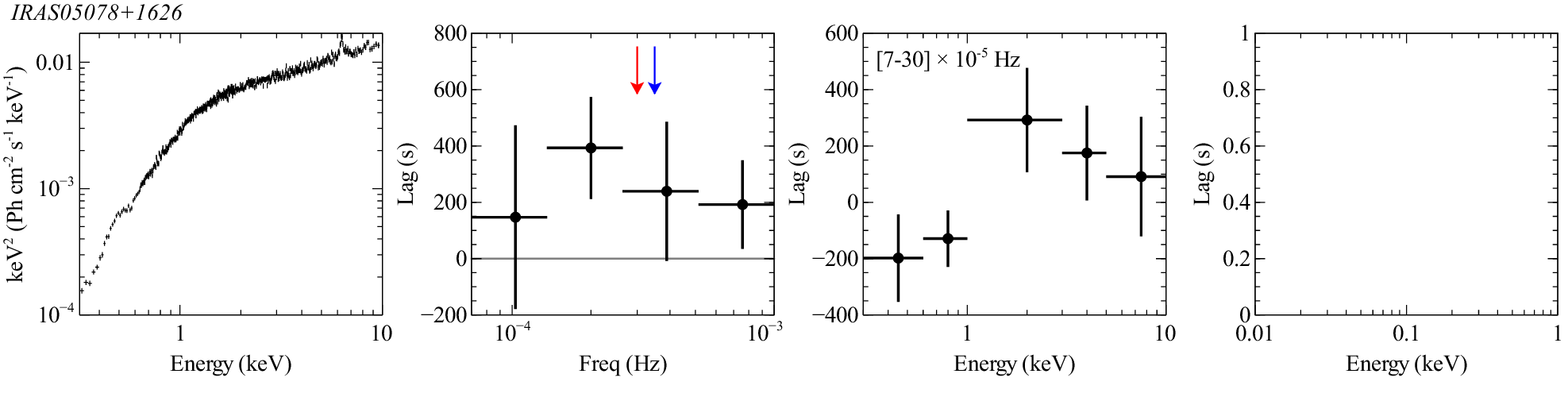}
\includegraphics[width=\textwidth]{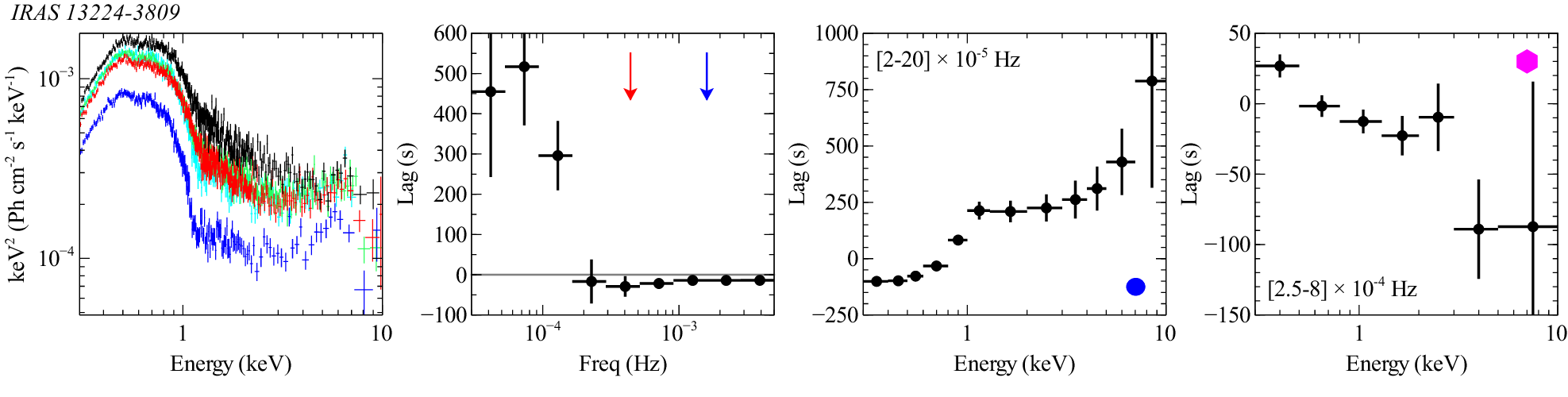}
\includegraphics[width=\textwidth]{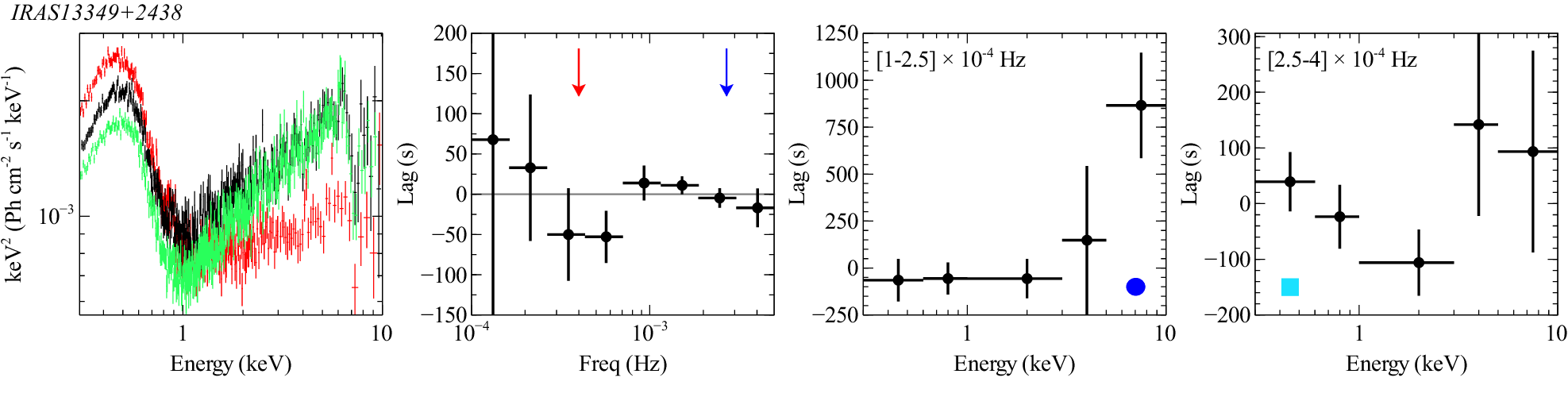}
\includegraphics[width=\textwidth]{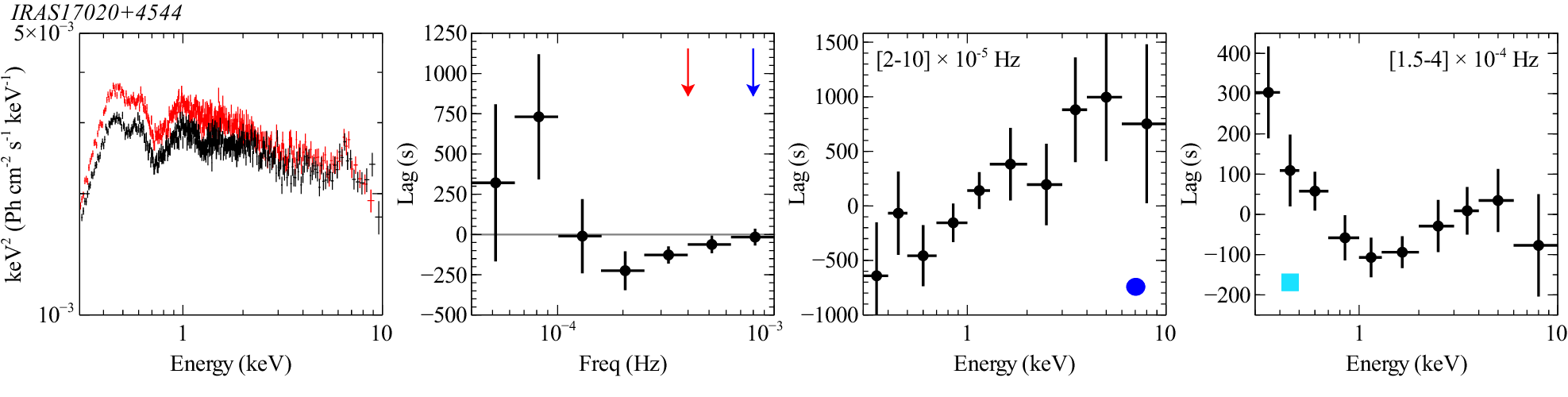}
\end{figure*}

\begin{figure*}
\includegraphics[width=\textwidth]{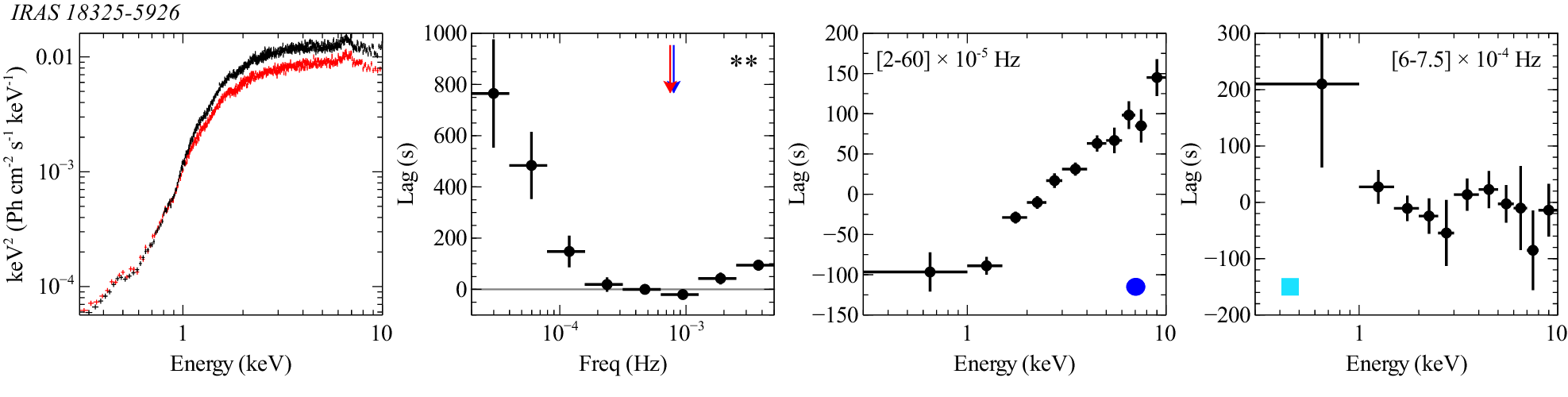}
\includegraphics[width=\textwidth]{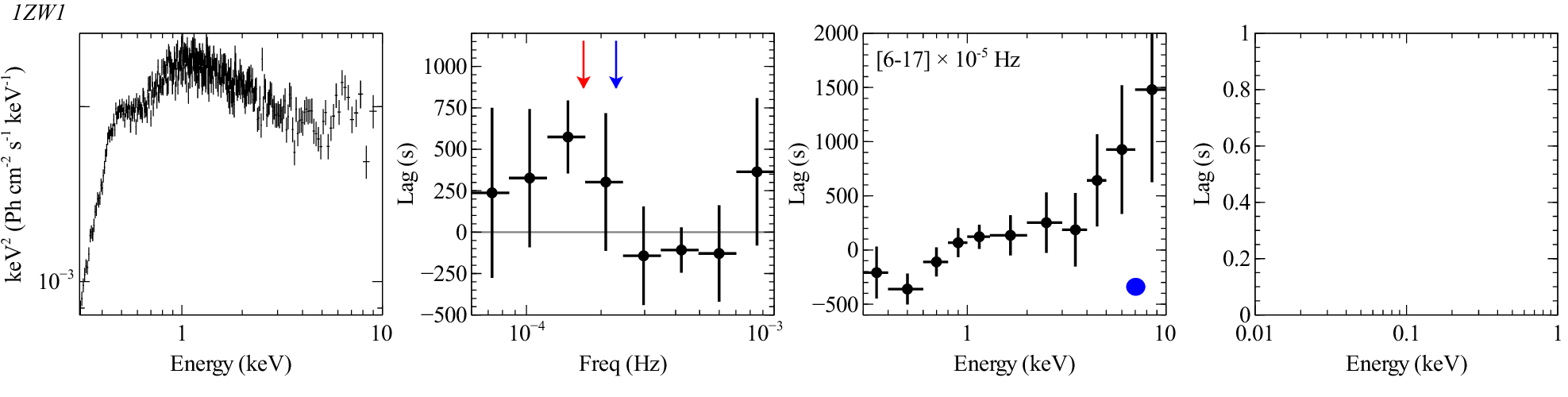}
\includegraphics[width=\textwidth]{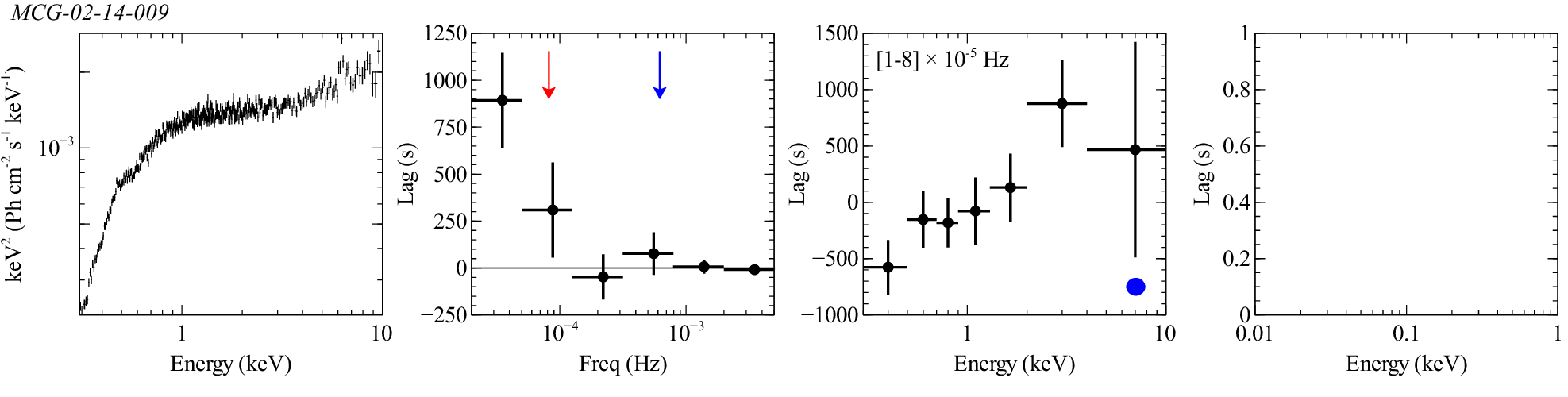}
\includegraphics[width=\textwidth]{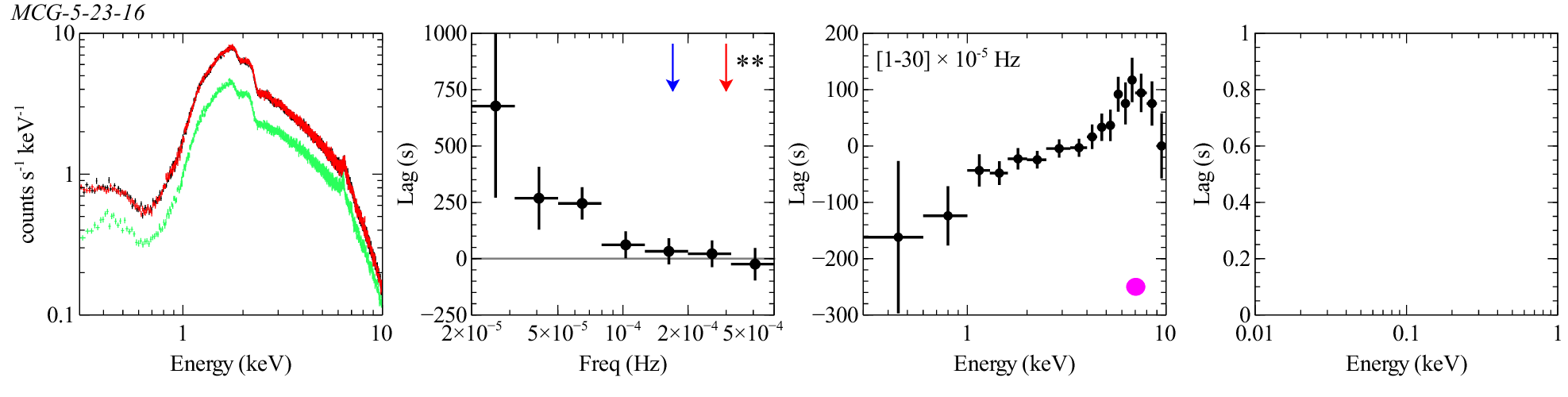}
\includegraphics[width=\textwidth]{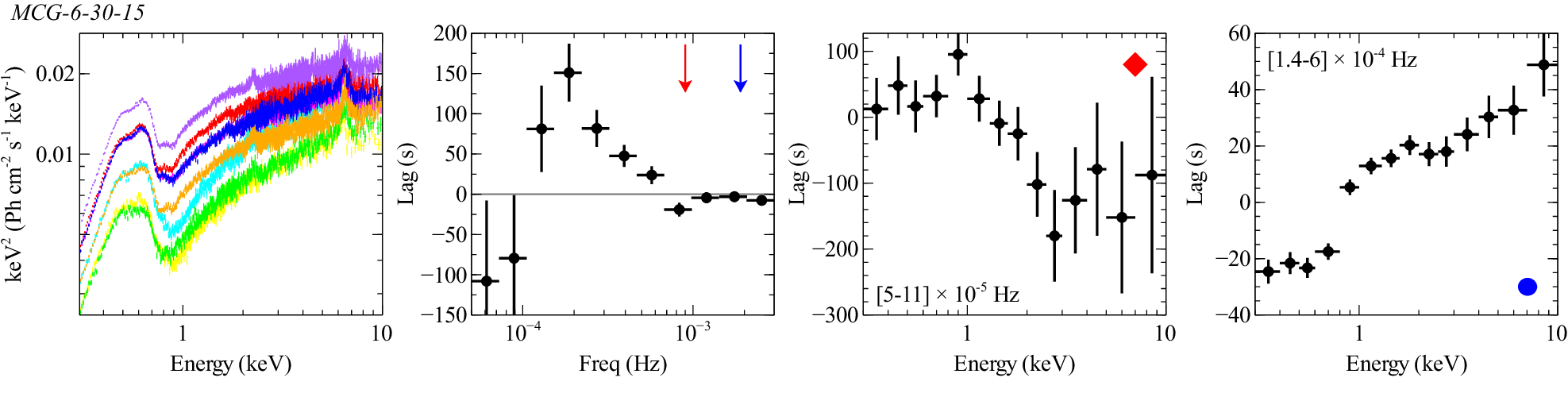}
\end{figure*}

\begin{figure*}
\includegraphics[width=\textwidth]{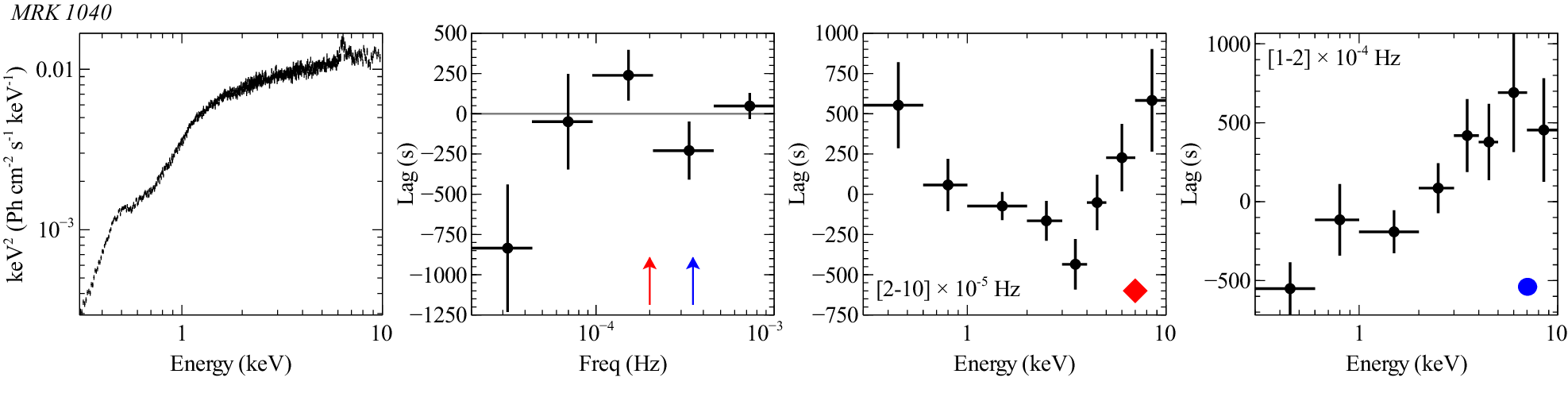}
\includegraphics[width=\textwidth]{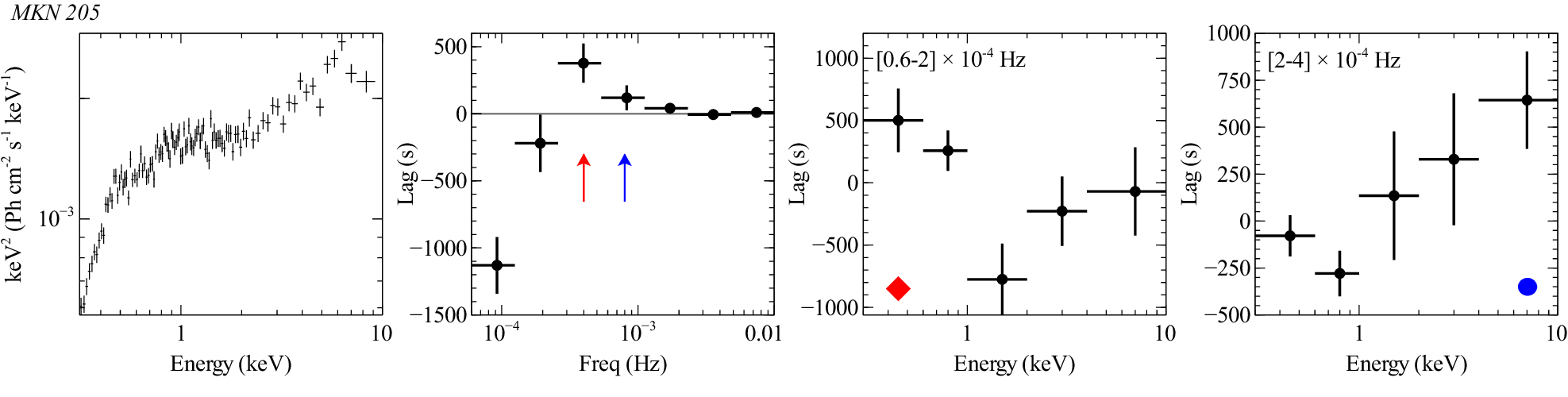}
\includegraphics[width=\textwidth]{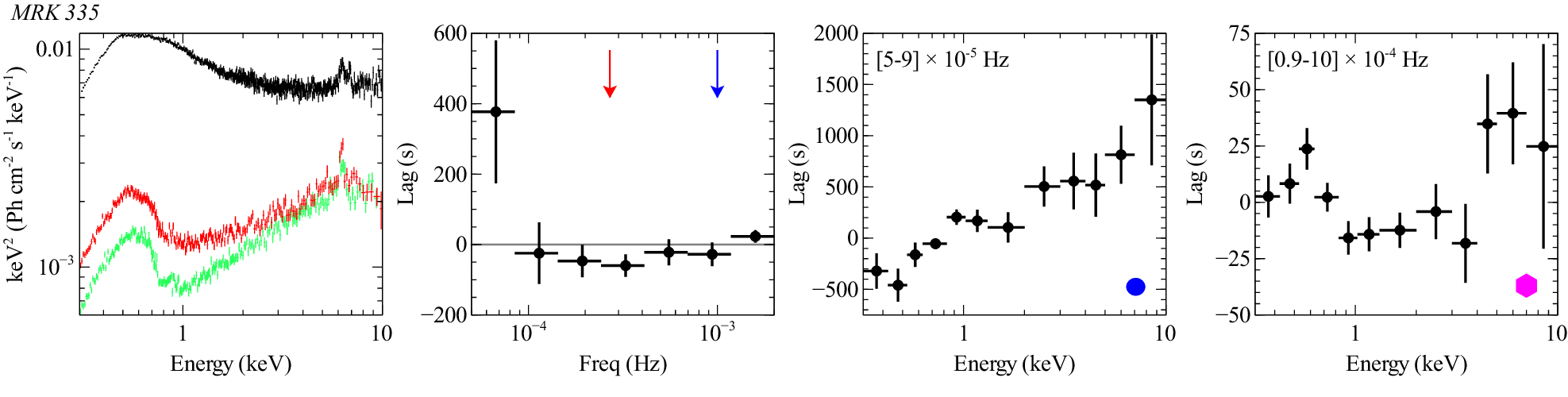}
\includegraphics[width=\textwidth]{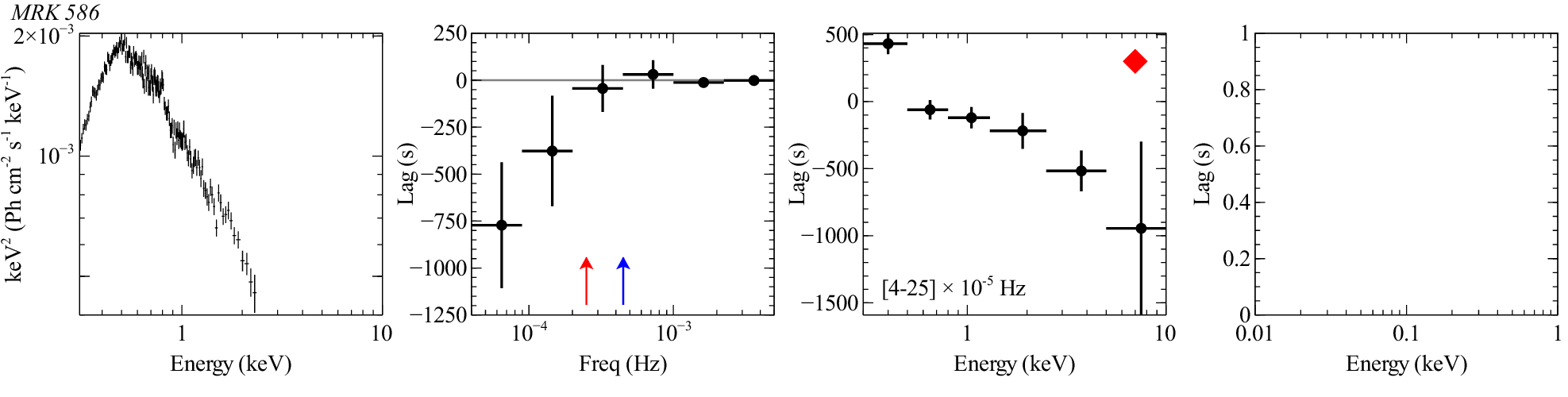}
\includegraphics[width=\textwidth]{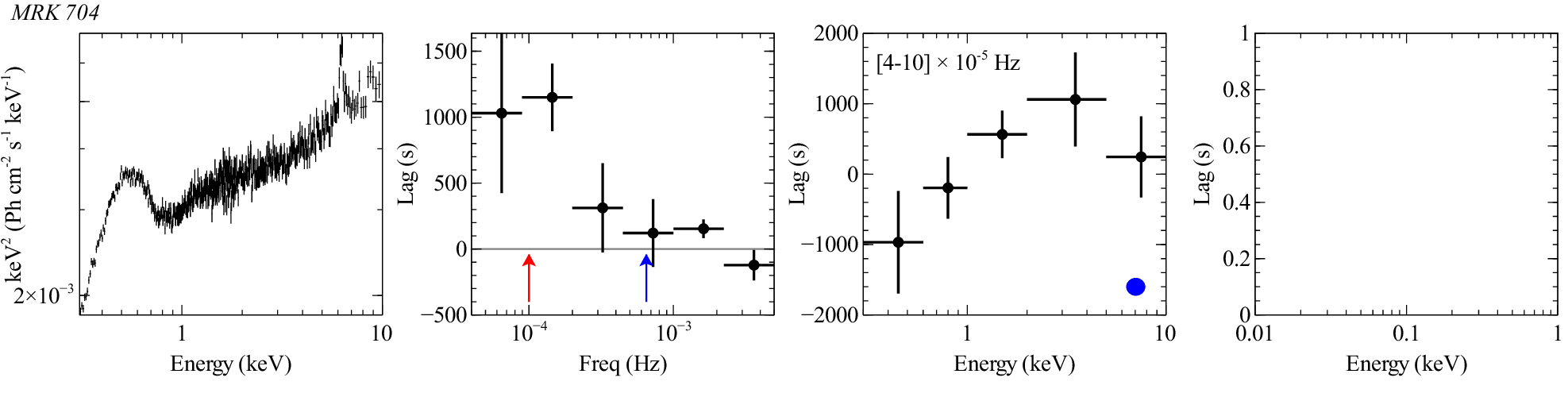}
\end{figure*}

\begin{figure*}
\includegraphics[width=\textwidth]{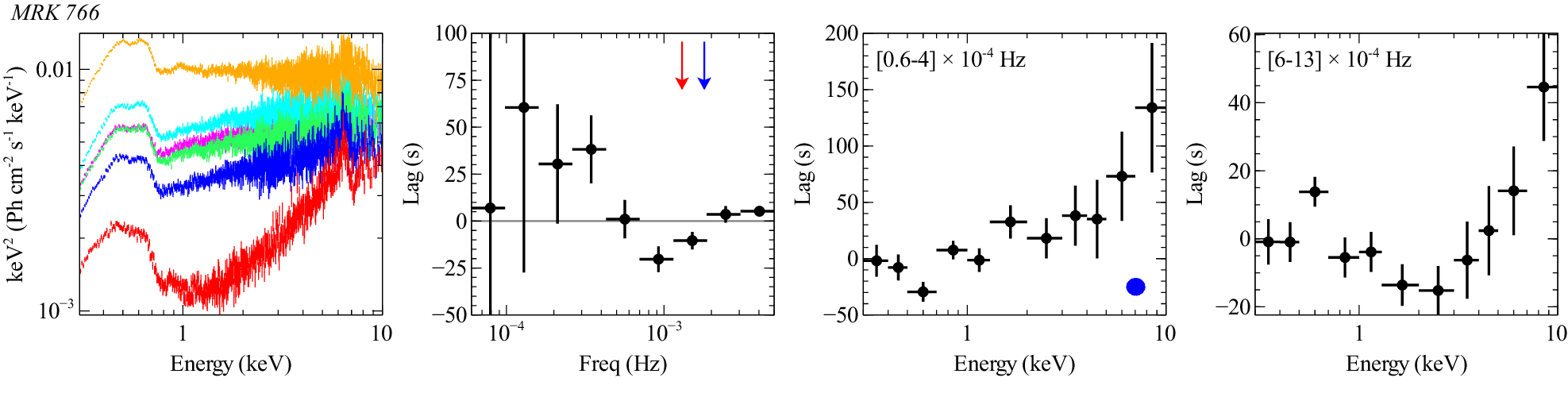}
\includegraphics[width=\textwidth]{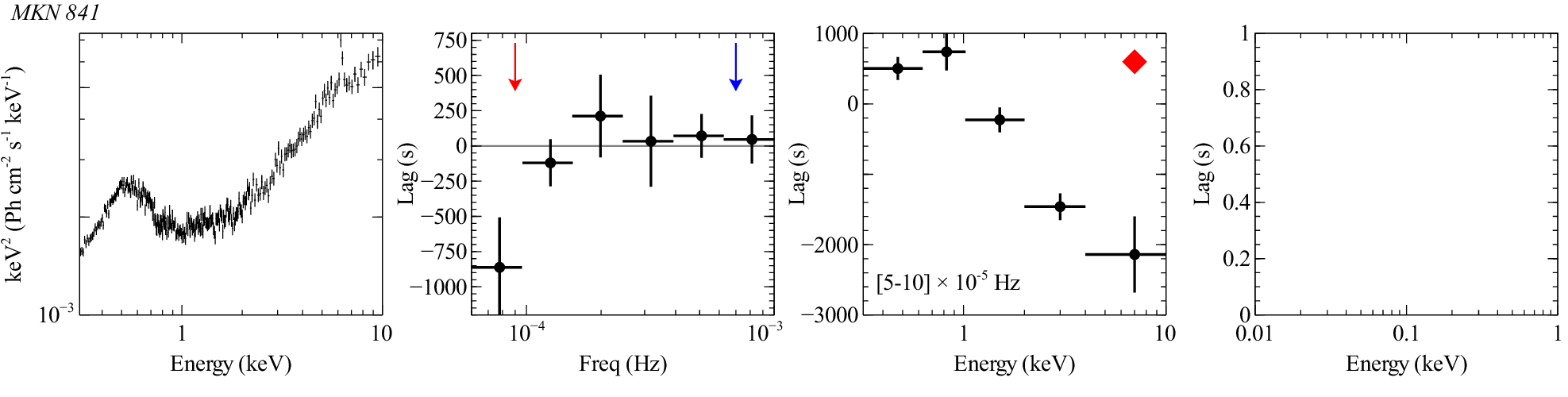}
\includegraphics[width=\textwidth]{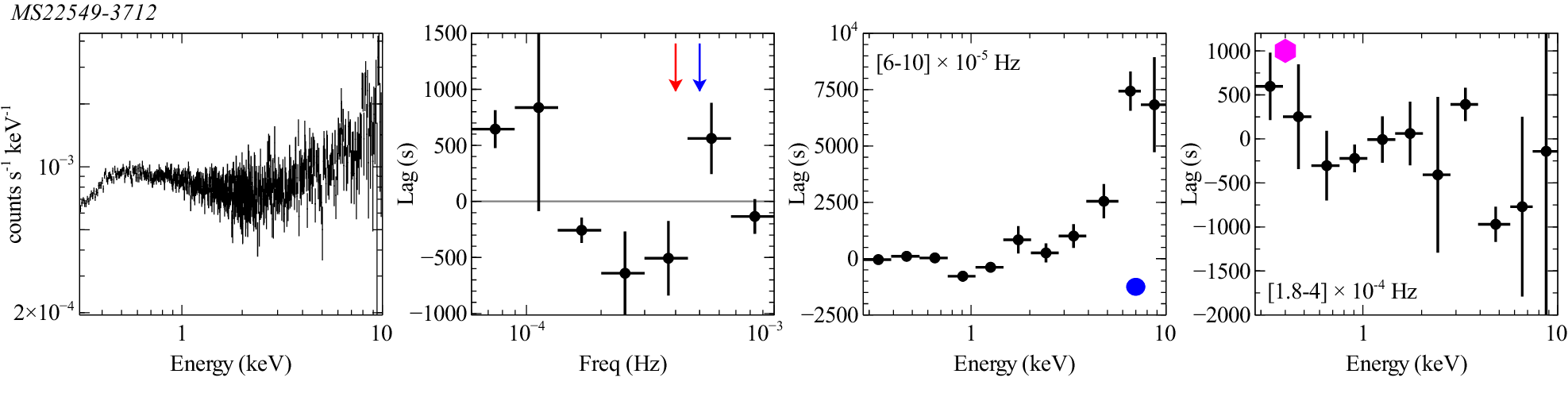}
\includegraphics[width=\textwidth]{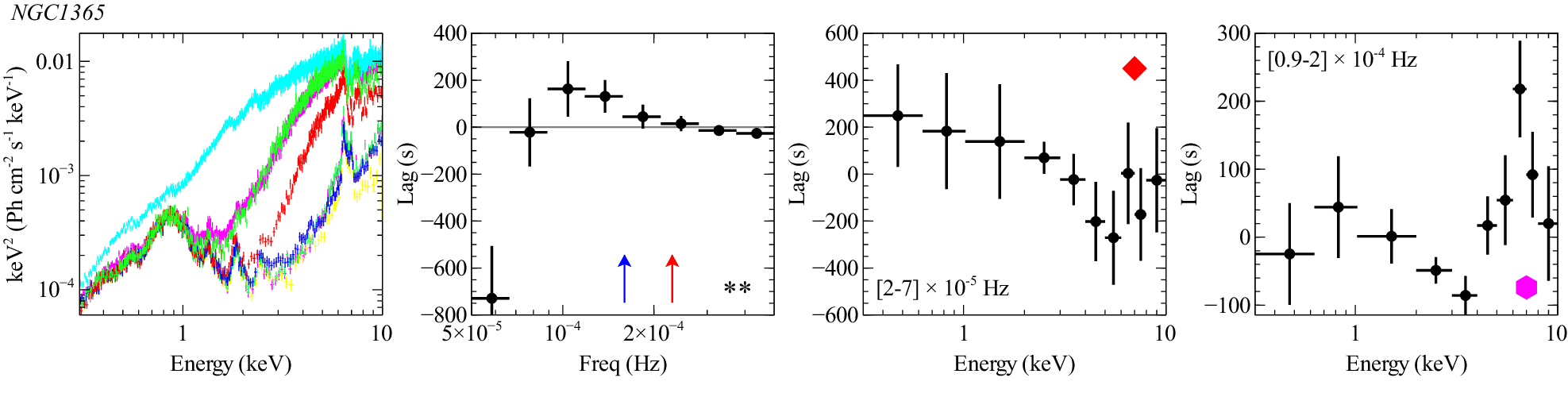}
\includegraphics[width=\textwidth]{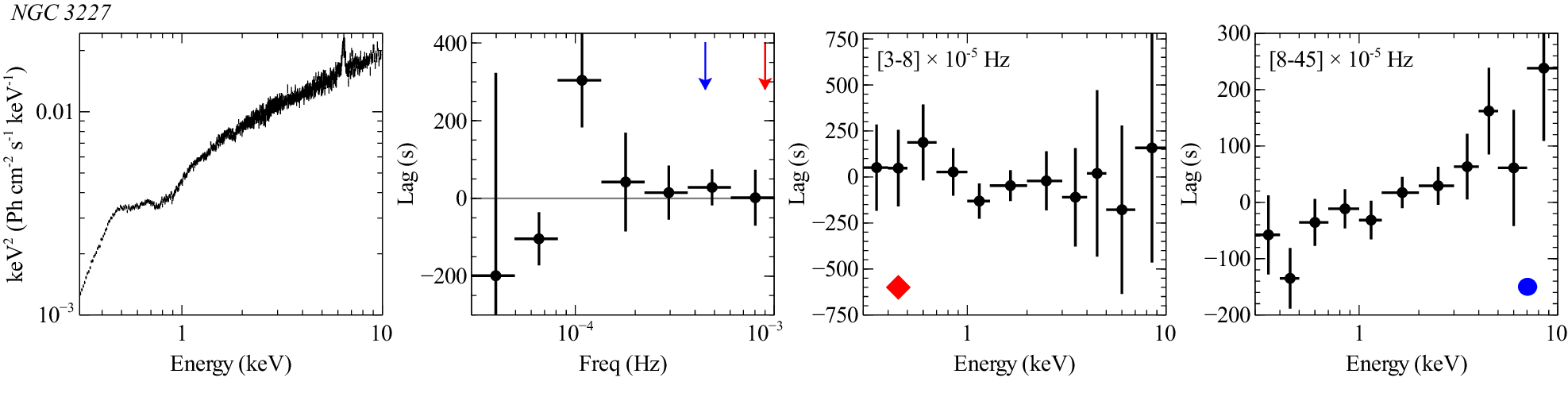}
\end{figure*}

\begin{figure*}
\includegraphics[width=\textwidth]{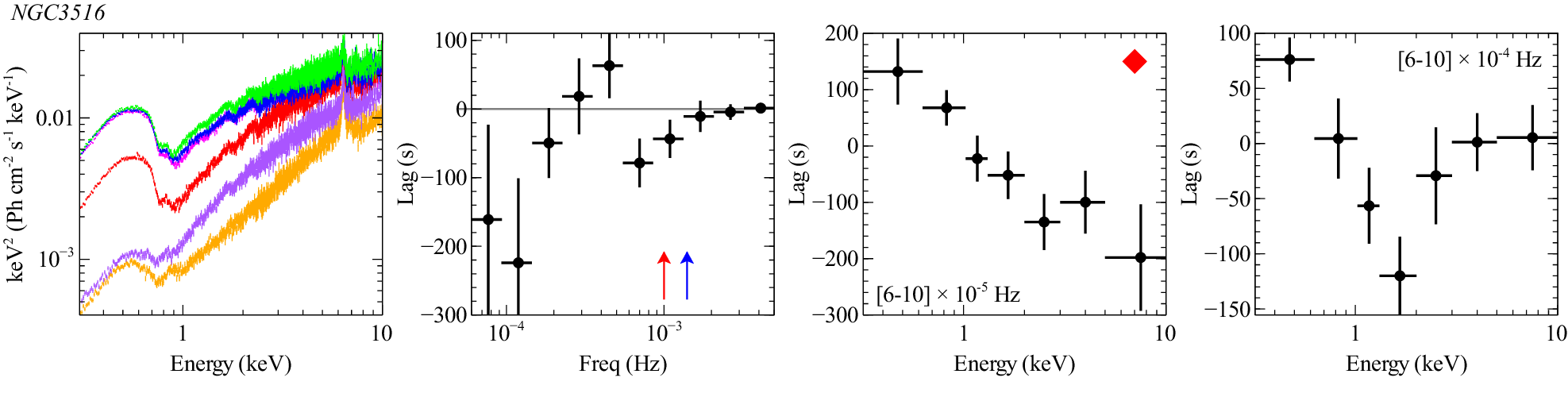}
\includegraphics[width=\textwidth]{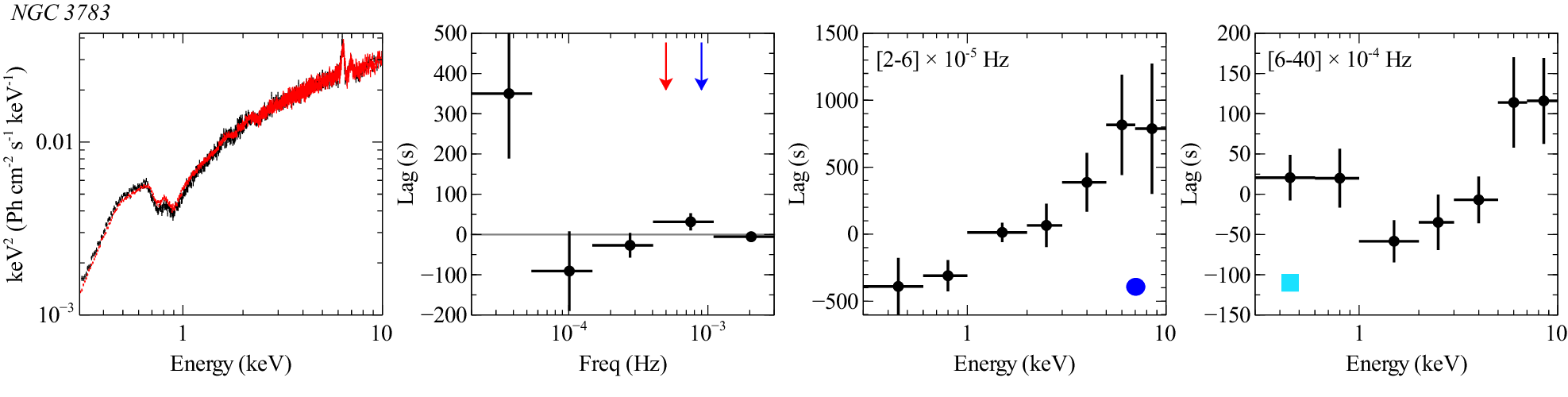}
\includegraphics[width=\textwidth]{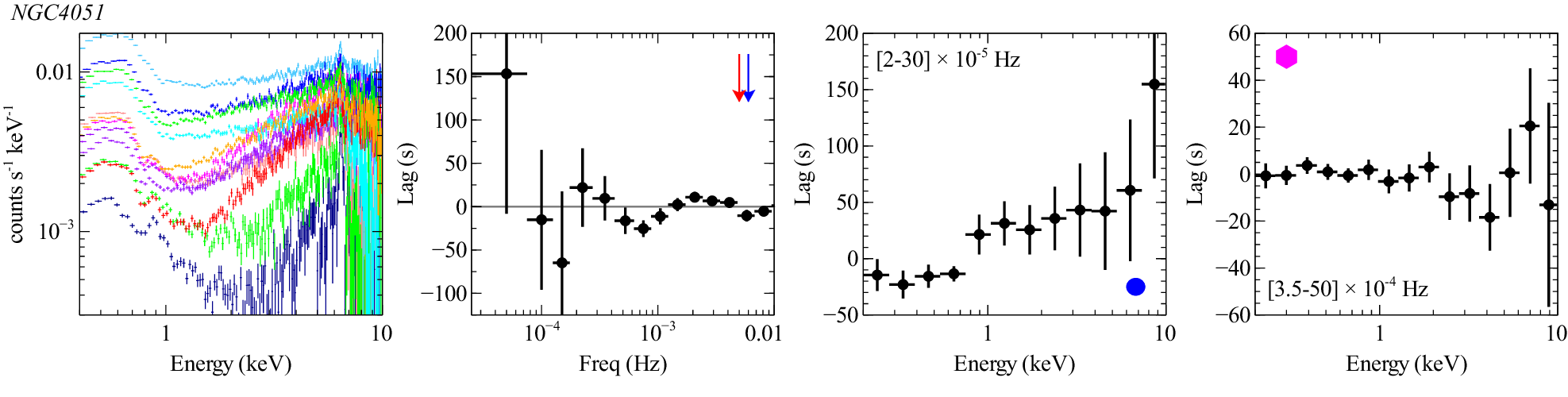}
\includegraphics[width=\textwidth]{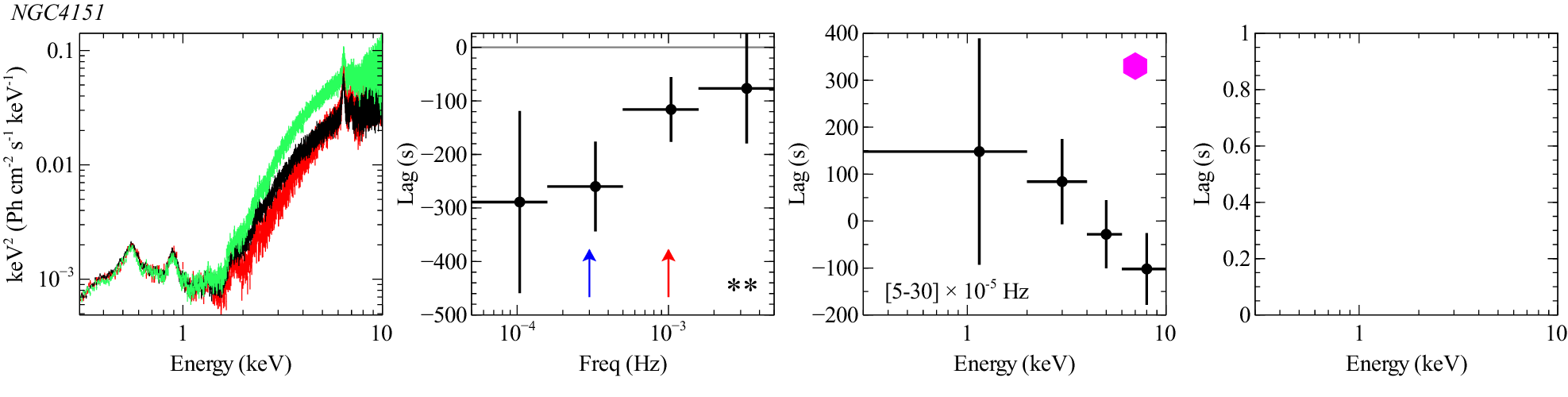}
\includegraphics[width=\textwidth]{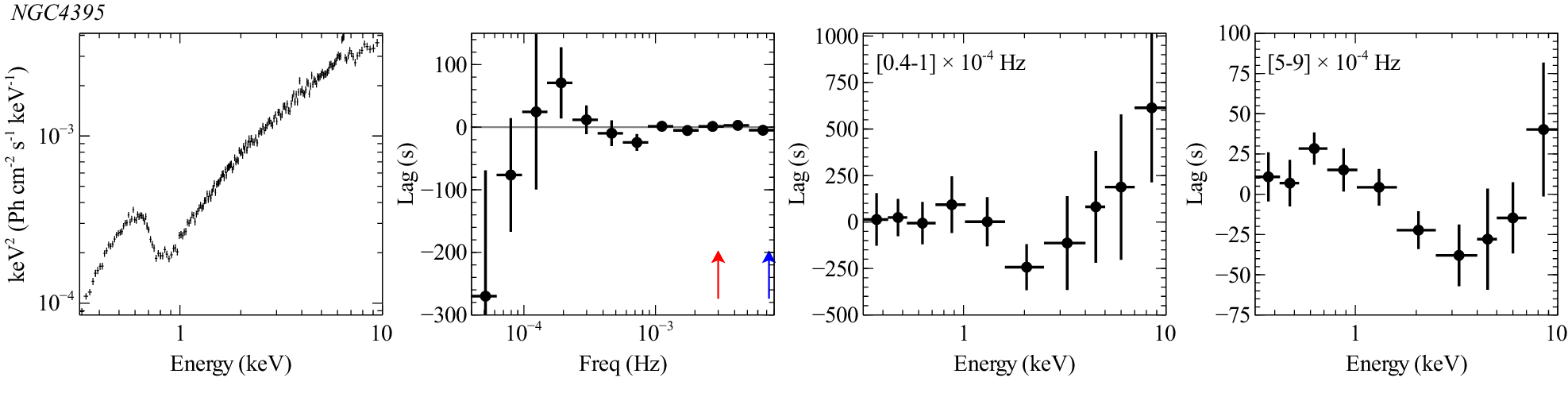}
\end{figure*}

\begin{figure*}
\includegraphics[width=\textwidth]{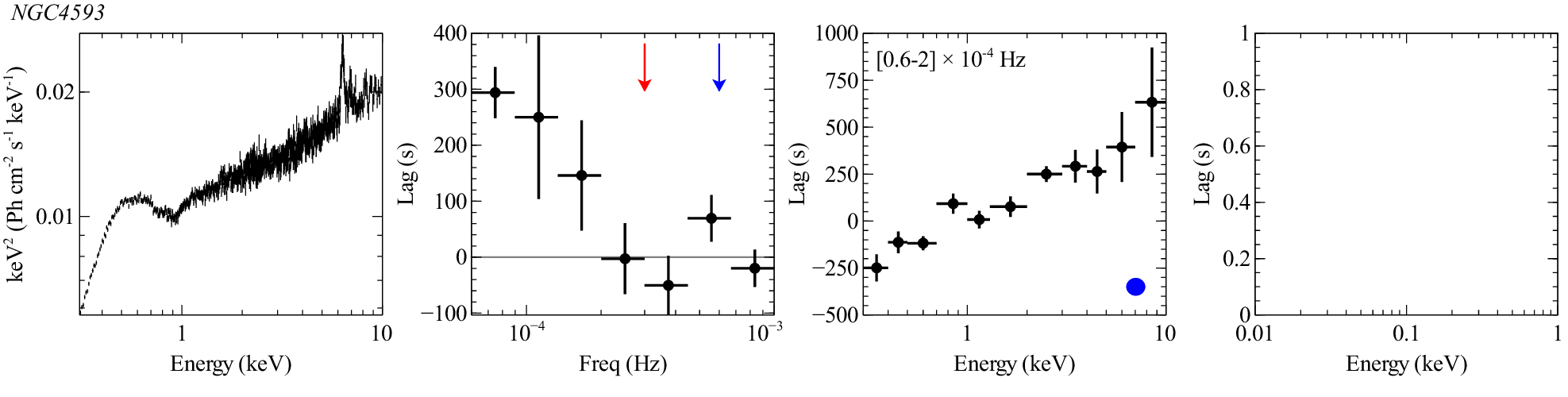}
\includegraphics[width=\textwidth]{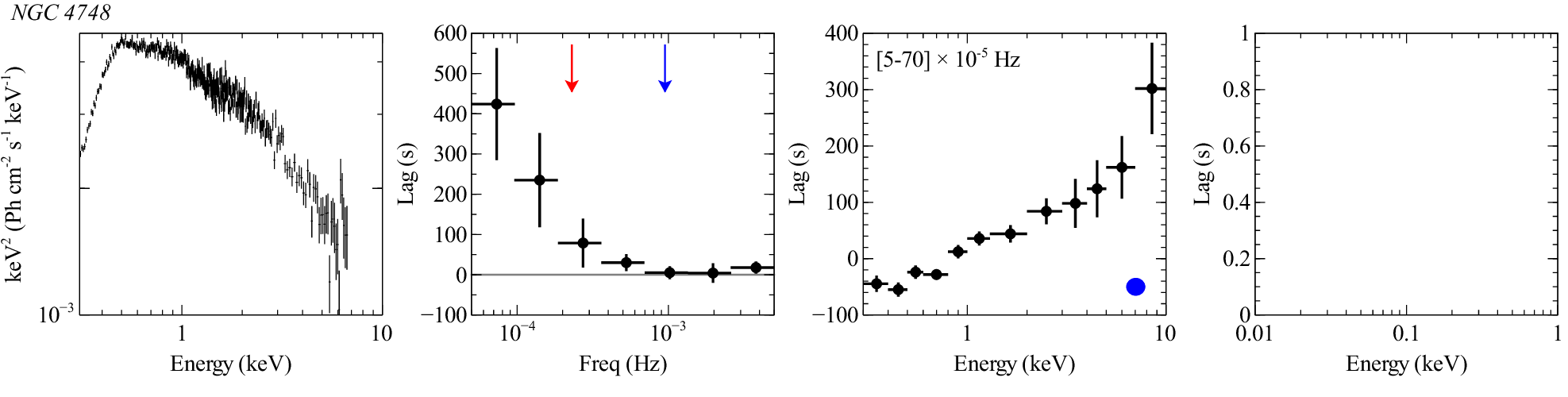}
\includegraphics[width=\textwidth]{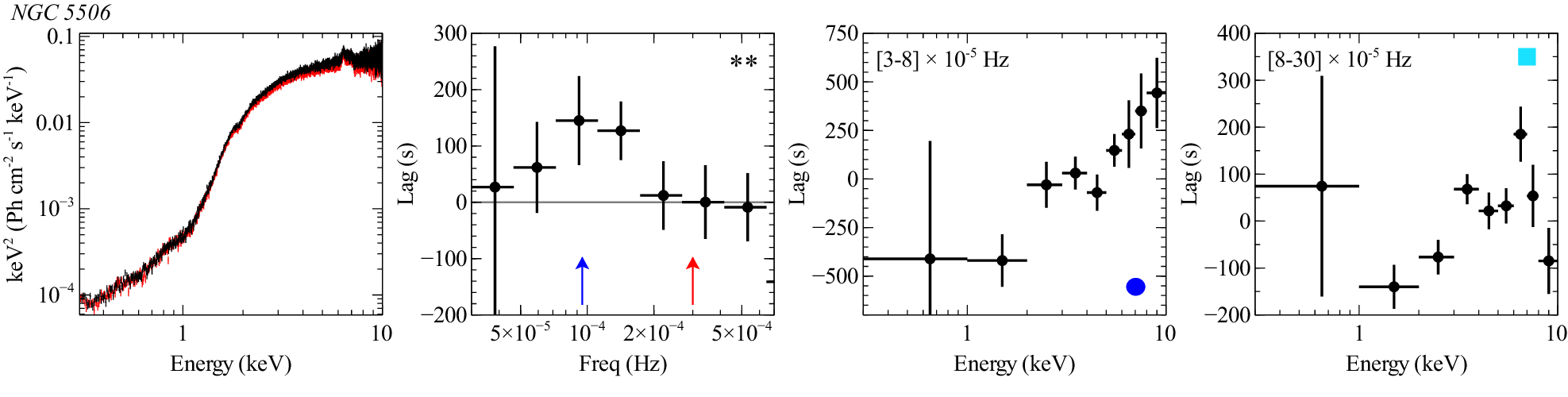}
\includegraphics[width=\textwidth]{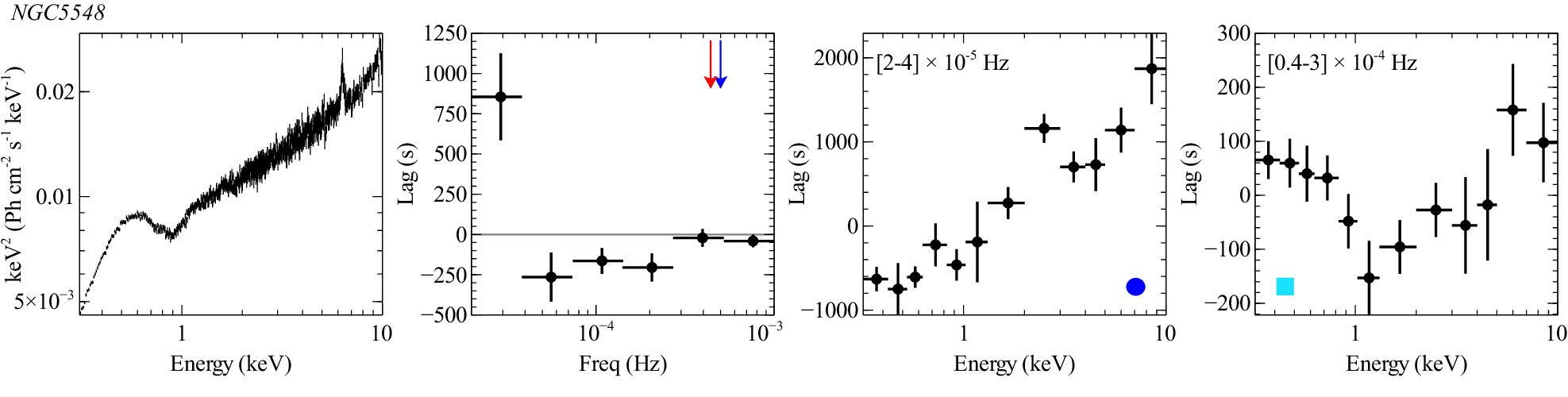}
\includegraphics[width=\textwidth]{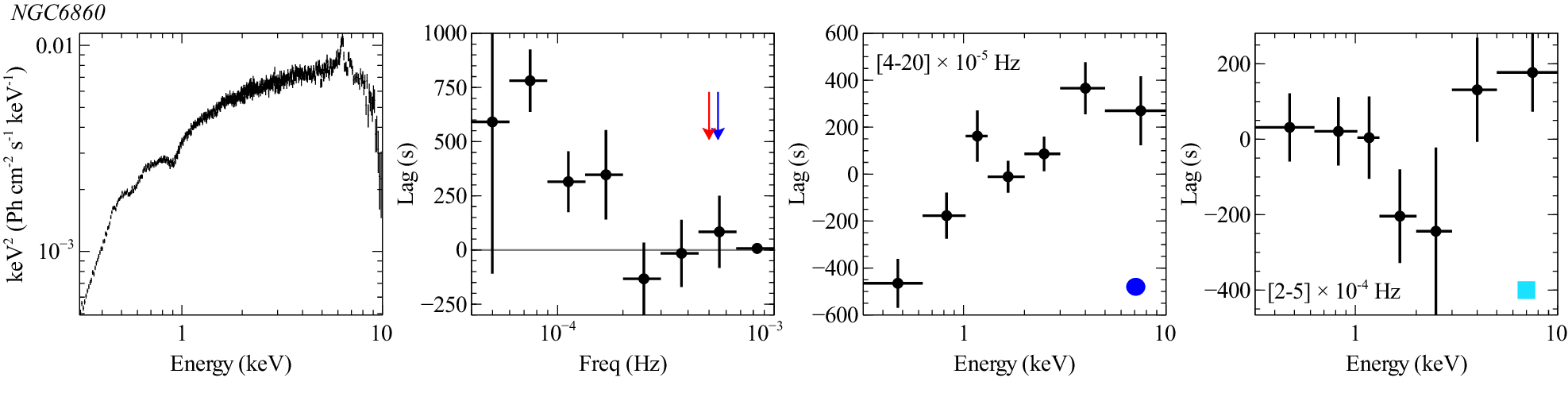}
\end{figure*}

\begin{figure*}
\includegraphics[width=\textwidth]{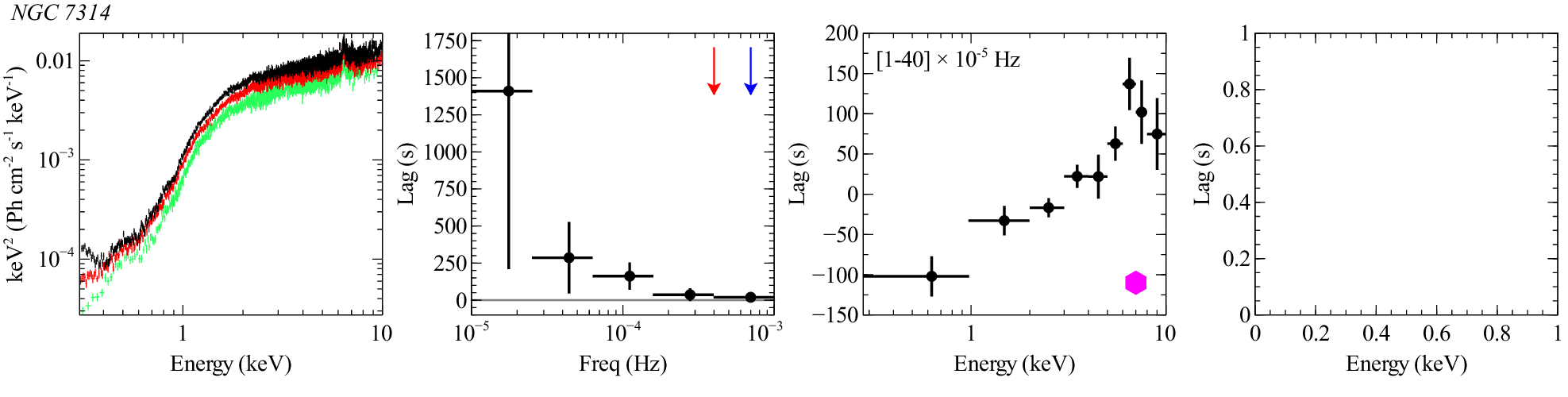}
\includegraphics[width=\textwidth]{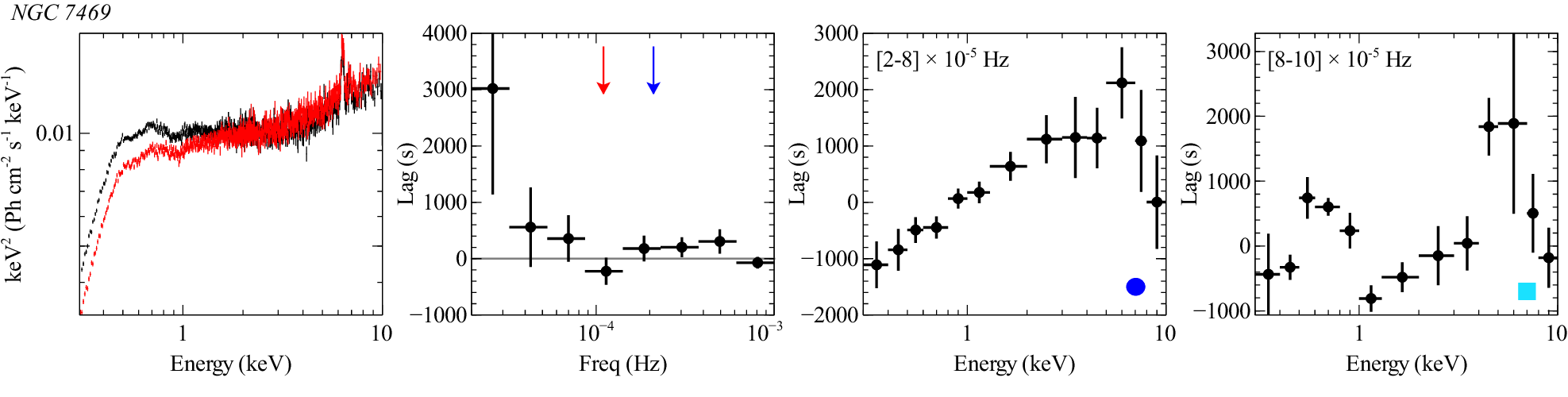}
\includegraphics[width=\textwidth]{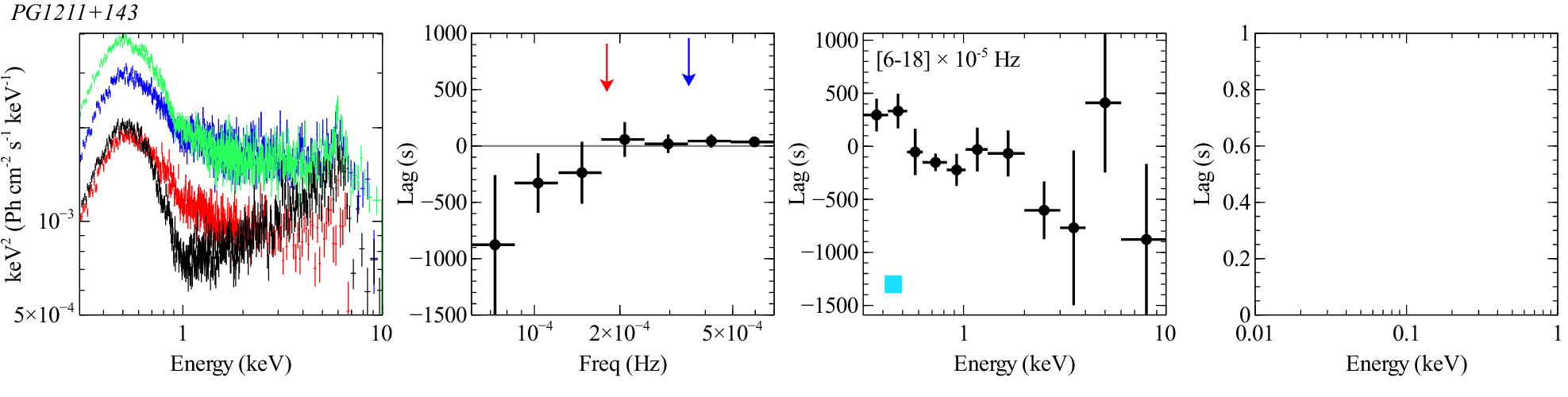}
\includegraphics[width=\textwidth]{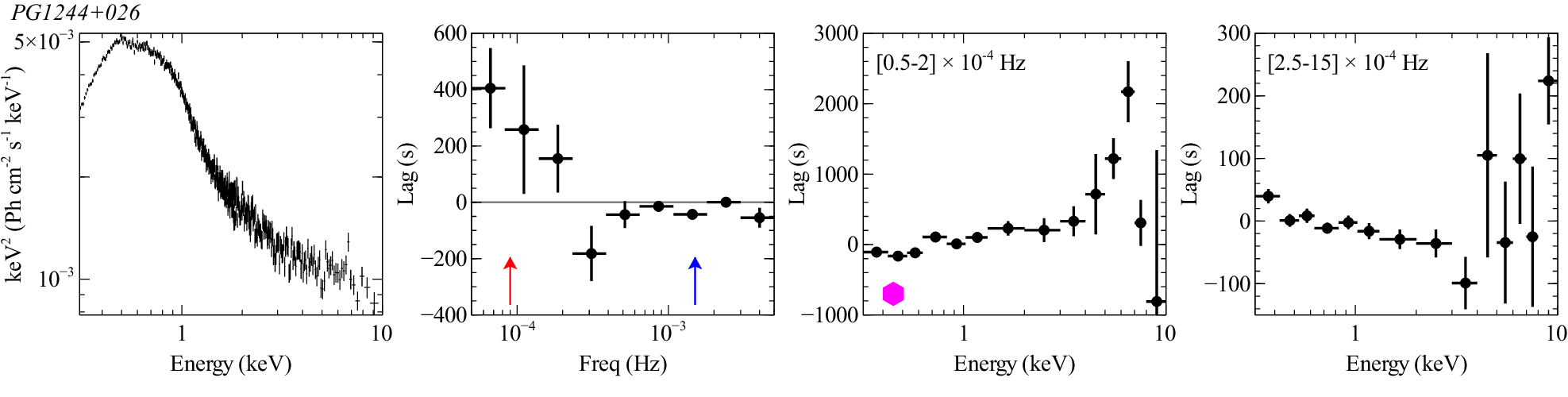}
\includegraphics[width=\textwidth]{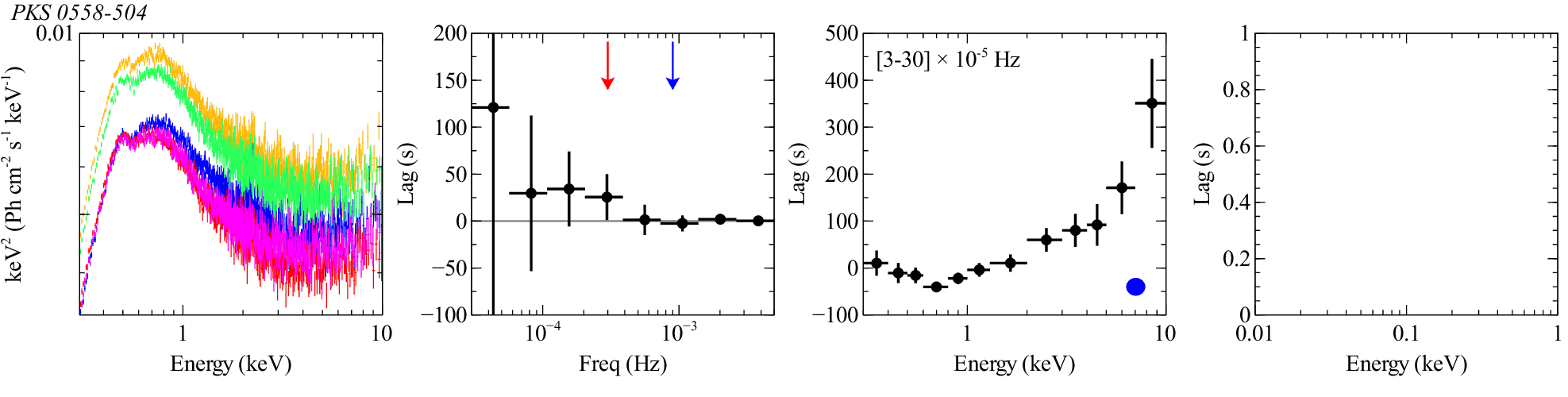}
\end{figure*}

\begin{figure*}
\includegraphics[width=\textwidth]{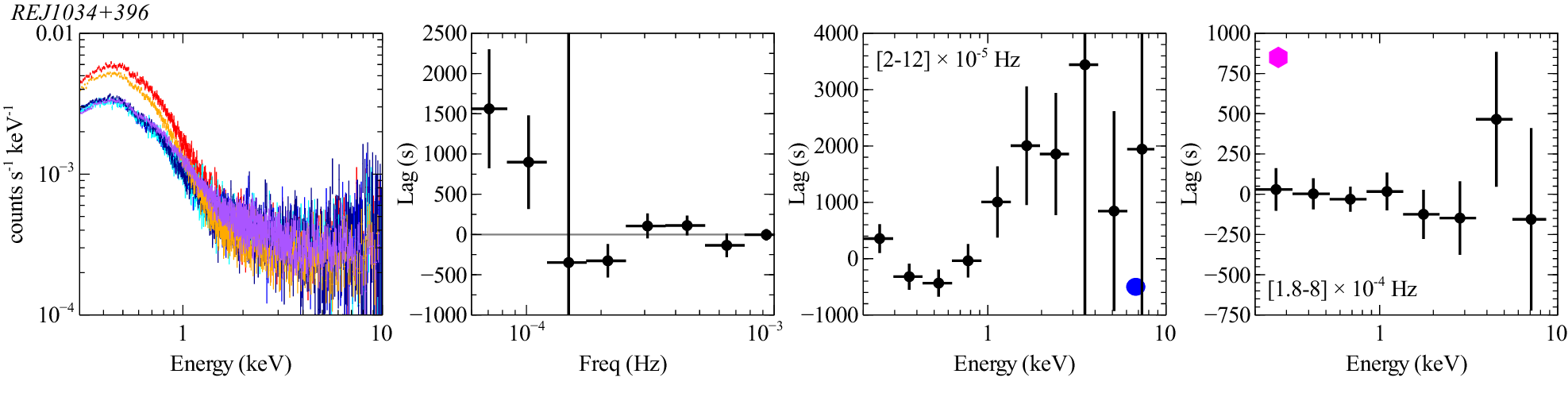}
\includegraphics[width=\textwidth]{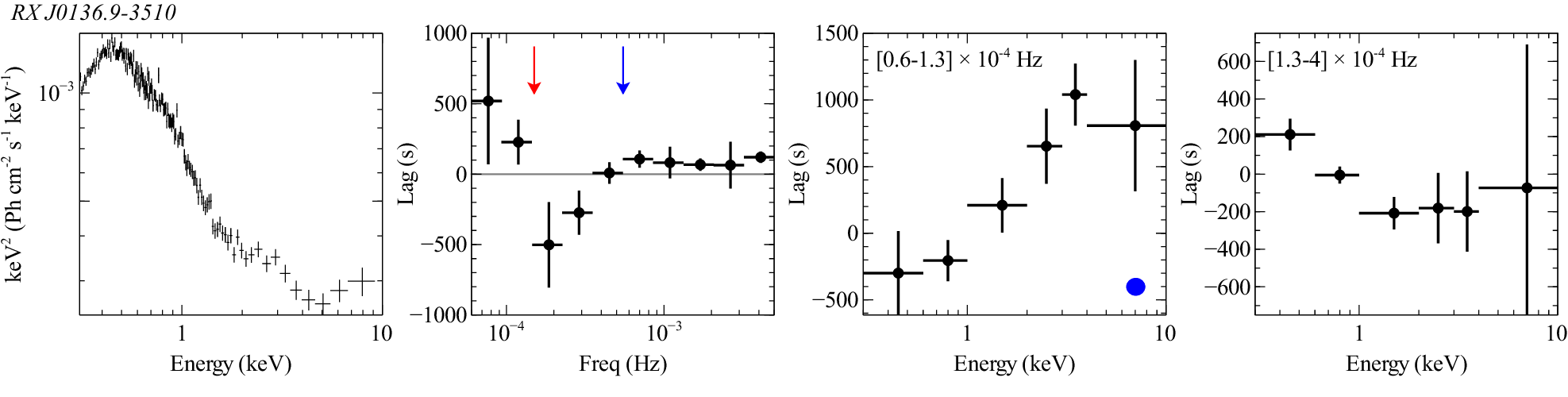}
\includegraphics[width=\textwidth]{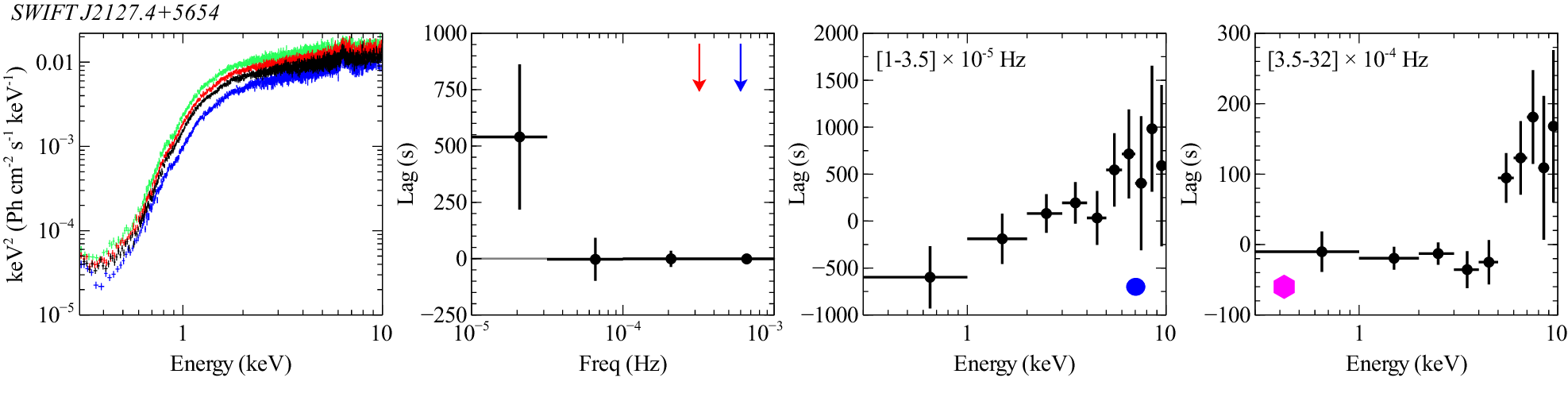}
\caption[X-ray time lag sample]{The X-ray time lags for the sample.  The leftmost panel shows the time-integrated fluxed energy spectrum for each of the individual observations used in this sample.  The next panel ({\em middle-left}) shows the lag-frequency spectrum between 0.3--1~keV and 1--4~keV. In a few cases, it was not possible to measure significant lag between the these bands because there was not much variability in the soft band. In those few cases (marked with two asterixes), we measured the lag between 2--4~keV and 4--7~keV.  The next panel ({\em middle-right}) shows the low-frequency lag-energy spectrum for the frequency range specified in the plot.  It is based on the low-frequency lags of the lag-frequency spectrum.  The rightmost panel shows the high-frequency lag-energy spectrum for sources whose lag-frequency spectrum changes at high frequencies.  The blue and red arrows on the lag-frequency spectra indicate the frequencies at which the lowest and highest energy bins of the lag-energy spectra become dominated by Poisson noise.  In nearly all cases, we do not probe frequencies above the red arrow. The symbols in the corners of the lag-energy spectra indicate generally what catagory they fall into: magenta hexagon = previously published iron~K reverberation; cyan square = iron K reverberation found through this work; blue circle = low-frequency hard lag; red diamond = low-frequency soft lag. Note that some sources with previously iron~K reverberation lags found reverberation at particular flux states (e.g. IRAS~13224-3809, NGC~4151, NGC~4051) or at the QPO frequenc (e.g. RE~J1034+396 and MS~22549-3712, and so we refer the interested reader to the iron~K discovery papers for a detailed analysis.}
\label{all_lags}
\end{figure*}

\end{document}